\documentclass{sig-alternate}
\usepackage{mathptmx} % This is Times font

\usepackage{fancyhdr}
\usepackage[normalem]{ulem}
\usepackage[hyphens]{url}
\usepackage[sort,nocompress]{cite}
\usepackage[final]{microtype}
\usepackage[keeplastbox]{flushend}

\usepackage{algorithmic}
\usepackage{bm}
\usepackage[braket]{qcircuit}
\usepackage{caption}
\usepackage{subcaption}
\usepackage{booktabs}
\usepackage{tabu}
\usepackage{adjustbox}

\usepackage[bookmarks=true,breaklinks=true,letterpaper=true,colorlinks,linkcolor=black,citecolor=blue,urlcolor=black]{hyperref}

\title{Optimized Quantum Compilation \\ for Near-Term Algorithms with OpenPulse}

\author{
Pranav Gokhale\thanks{pranavgokhale@uchicago.edu} \\ University of Chicago \and
Ali Javadi-Abhari \\ IBM
\and Nathan Earnest \\ IBM
\and Yunong Shi \\ University of Chicago \and Frederic T. Chong \\ University of Chicago \\}

\begin{document}
\maketitle
\pagestyle{plain}

%%%%%% -- PAPER CONTENT STARTS-- %%%%%%%%
\begin{abstract}
Quantum computers are traditionally operated by programmers at the granularity of a gate-based instruction set. However, the actual device-level control of a quantum computer is performed via analog pulses. We introduce a compiler that exploits direct control at this microarchitectural level to achieve significant improvements for quantum programs. Unlike quantum optimal control, our approach is bootstrapped from existing gate calibrations and the resulting pulses are simple. Our techniques are applicable to any quantum computer and realizable on current devices. We validate our techniques with millions of experimental shots on IBM quantum computers, controlled via the OpenPulse control interface. For representative benchmarks, our pulse control techniques achieve both 1.6x lower error rates and 2x faster execution time, relative to standard gate-based compilation. These improvements are critical in the near-term era of quantum computing, which is bottlenecked by error rates and qubit lifetimes.
\end{abstract}

\section{Introduction}
The present era of quantum computing is characterized by the emergence of quantum computers with dozens of qubits, as well as new algorithms that have innate noise resilience and modest qubit requirements. There are promising indications that near-term devices could be used to accelerate or outright-enable solutions to problems in domains ranging from molecular chemistry \cite{mcardle2018quantum} to combinatorial optimization \cite{farhi2014quantum} to adversarial machine learning \cite{anschuetz2019near}. To realize these practical applications on noisy hardware, it is critical to optimize across the full stack, from algorithm to device.

Standard quantum compilers operate at the level of gates. However, the lowest-level of quantum control is through analog pulses. Pulse optimization has shown promise in previous quantum optimal control (QOC) work \cite{song2019generation, omran2019generation}, but we found that noisy experimental systems are not ready for compilation via QOC approaches. This is because QOC requires an extremely accurate model of the machine, i.e. its Hamiltonian. Hamiltonians are difficult to measure experimentally and moreover, they drift significantly between daily recalibrations. Experimental QOC papers incur significant pre-execution calibration overhead to address this issue. By contrast, we propose a technique that is bootstrapped purely from daily calibrations that are already performed for the standard set of basis gates. The resulting pulses form our \textbf{augmented basis gate set}. These pulses are extremely simple, which reduces control error and also preserves intuition about underlying operations, unlike QOC. This technique leads to optimized programs, with mean 1.6x error reduction and 2x speedup for near-term algorithms.

\begin{figure}
    \centering
    \includegraphics[width=0.45\textwidth]{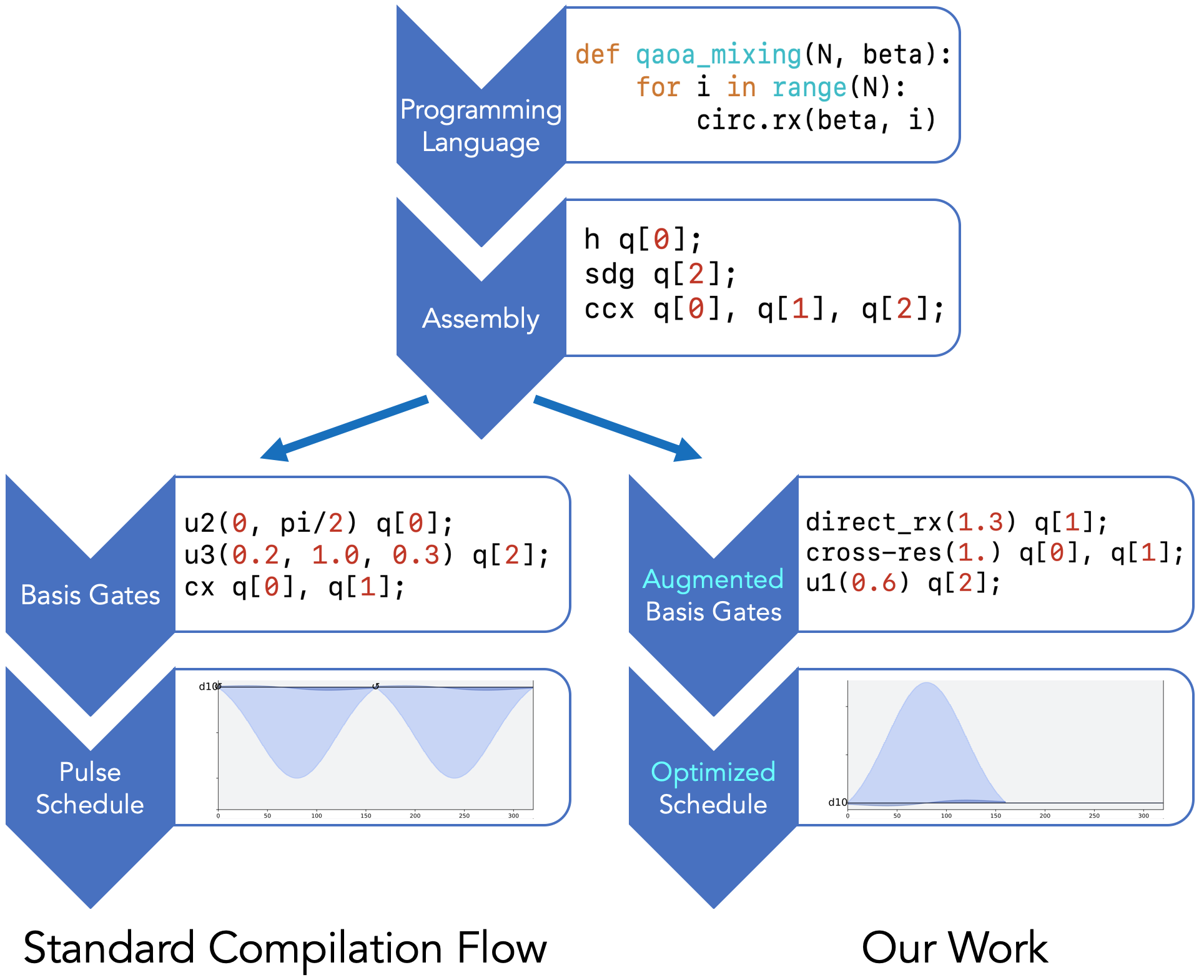} 
    \caption{Like classical programs, quantum programs undergo a compilation process from high-level programming language to assembly. However, unlike the classical setting, quantum hardware is controlled via analog pulses. In our work, we optimize the underlying pulse schedule by augmenting the set \textit{basis gates} to match hardware. Our compiler automatically optimizes user code, which therefore remains hardware-agnostic.}
    \label{fig:compilation_flow}
\end{figure}

We emphasize the generality of our approach and our compiler, which can target any underlying quantum hardware. We demonstrate our results via OpenPulse \cite{mckay2018qiskit, alex2020qiskit}, an interface for pulse-level control. In particular, our work is the first experimental demonstration of OpenPulse for optimized compilation of quantum programs (one prior paper used OpenPulse for noise extrapolation \cite{garmon2019benchmarking}). We executed pulse schedules on IBM's 20-qubit Almaden quantum computer, accessible through the cloud via the IBM Q Experience~\cite{ibmqx}. Our experience-building spanned over 11.4 million experimental shots, 4 million of which are explicitly presented here as concrete research outcomes. Our results indicate that  pulse-level control significantly extends the computational capacity of quantum computers. Our techniques are realizable immediately on existing OpenPulse-compatible devices. To this end, all of our code and notebooks are available on Github \cite{anonymized_repo}.

We begin with background on quantum computing in Section~\ref{sec:background}. Next, Section~\ref{sec:compiler_flow} presents an overview of standard quantum compilers and our compiler design (depicted in Figure~\ref{fig:compilation_flow}). Sections~\ref{sec:direct_drive}--~\ref{sec:qudit_operations} describe four key optimizations in our compiler, all of which are enabled by pulse-level control:

\begin{enumerate}
    \item \textbf{Direct Rotations} (Section~\ref{sec:direct_drive}). Access to pulse-level control allows us to implement any single-qubit operation \textit{directly} with high fidelity, circumventing inefficiencies from standard compilation.
    \item \textbf{Cross-Gate Pulse Cancellation} (Section~\ref{sec:cross_pulse_cancellation}). Although gates have the illusion of atomicity, the true atomic units are pulses. Our compiler creates new cancellation optimizations that are otherwise invisible.
    \item \textbf{Two-Qubit Operation Decompositions} (Section~\ref{sec:cross_resonance_optimizations}). We recompile important near-term algorithm primitives for two-qubit operations directly down to the two-qubit interactions that hardware actually implements.
    \item \textbf{Qudit Operations} (Section~\ref{sec:qudit_operations}). Quantum systems have infinite energy levels. Pulse control enables d-level \textit{qudit} operations, beyond the 2-level qubit subspace.
\end{enumerate}

Section~\ref{sec:results} presents results from application of these techniques to full algorithms. We conclude in Section~\ref{sec:conclusion}.
\section{Background} \label{sec:background}

We assume some familiarity with the fundamentals of quantum computing. Here we provide a brief review and expand on elements relevant to our work.

\subsection{The Qubit}

\begin{figure}
    \centering
    \includegraphics[width=0.28\textwidth]{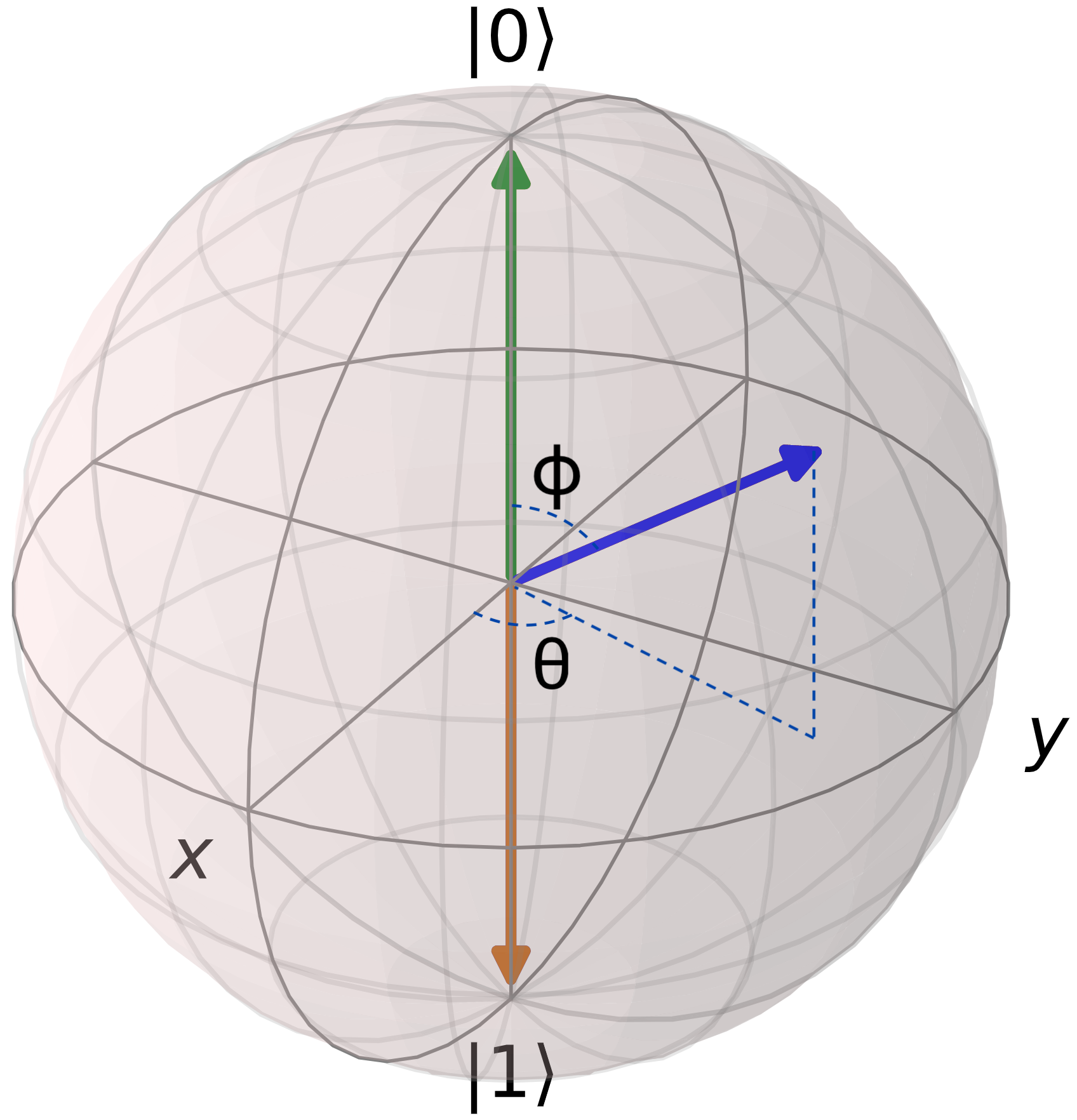}
    \caption{The points on the Bloch sphere correspond one-to-one with possible qubit states. The green and brown states correspond to $\ket{0}$ and $\ket{1}$ respectively. The blue state is in a superposition described by latitude and longitude angles.}
    \label{fig:bloch_states}
\end{figure}

The core unit involved in quantum computation is the qubit (quantum bit). Unlike a classical bit which is either 0 or 1, a qubit can occupy any \textit{superposition} between the two states, which are now denoted $\ket{0}$ and $\ket{1}$. The Bloch sphere, depicted in Figure~\ref{fig:bloch_states}, is a useful visual representation of the possible states of a qubit. The North Pole is the $\ket{0}$ state, and the South Pole is the $\ket{1}$ state. The state of a qubit can be parametrized by two angles, latitude and longitude. Upon measurement, a qubit \textit{collapses} to either the $\ket{0}$ or $\ket{1}$ state, with probabilities dependent only on the latitude.

\subsection{Quantum Gates}
The set of valid single-qubit gates correspond to rotations around the Bloch sphere. Arbitrary such rotations are typically decomposed into $R_x(\theta)$ and $R_z(\theta)$ rotations around the X and Z axes respectively, which are universal for single-qubit rotations \cite{nielsen2002quantum}. A prominent single-qubit gate is the $X = R_x(180^{\circ})$ gate, which in Figure~\ref{fig:bloch_states} would rotate the green $\ket{0}$ state $180^{\circ}$ around the X-axis to the $\ket{1}$ and vice versa. Thus, the $X$ operation implements the NOT gate.

The set of possible multiple-qubit operations is much richer than the set of single-qubit gates, and lacks a clear visualization on the Bloch sphere. Remarkably however, any multiple-qubit operation can be decomposed into single-qubit rotations + an \textit{entangling} gate such as CNOT \cite{nielsen2002quantum}. The CNOT gate acts on a control and target qubit, and it applies $X$ to the target iff the control is $\ket{1}$. In part because the CNOT gate is easy to understand, most quantum programs are expressed in terms of it. By implementing a small set of gates: single qubit rotations + CNOT, a quantum computer is universal. Accordingly, quantum computers are generally designed with this interface in mind. However, the lowest level of hardware control is performed by microwave pulses. Foreshadowing the main message of our paper: this pulse-backed layer actually provides a richer and ``overcomplete'' set of gates that outperforms the standard interface for quantum programs.

\subsection{Gate Calibration} \label{subsec:calibration}
To implement this standard interface of universal gates, quantum computers are routinely calibrated to account for continuous drift in the experimental setting \cite{murali2019noise, tannu2019not}. As a concrete example, for superconducting devices, an $R_x(90^\circ)$ gate is calibrated by performing a Rabi experiment \cite{majer2007coupling, galperin2005rabi, ashhab2006rabi} that determines the necessary underlying pulses. An additional DRAG \cite{motzoi2009simple, gambetta2011analytic, motzoi2013improving} calibration fine-tunes the $R_x(90^\circ)$ gate by cancelling out stray components. Calibrations in a similar spirit are also performed for the two-qubit gate(s). An interesting feature of the two-qubit gate calibrations is that they have the side effect of also calibrating $R_x(180^\circ)$ pulses on each qubit. We exploit this free calibration in Section~\ref{sec:direct_drive}. Typically, $R_Z(\theta)$ rotation gates do not require calibration because they are implemented in software, as described in Section~\ref{sec:direct_drive}.

\subsection{Experimental Setup}
Our experiments were performed on IBM's Almaden, a 20 qubit device \cite{almaden}. Almaden is the first cloud-accessible OpenPulse device. It comprises 20 transmon qubits, with mean $T_1$ and $T_2$ coherence lifetimes of 94 and 88 $\mu$s respectively. The mean single-qubit and two-qubit  (CNOT) error  rates  are  0.14\%  and  1.78\%.  The  mean  measurement (readout) error was 3.8\%, though we used measurement error mitigation \cite{maciejewski2019mitigation, chen2019detector} to correct for biased measurement errors.

As of December 2019, IBM's publicly cloud-accessible OpenPulse device is the new Armonk device \cite{armonk}, which we  used for  the  most  recent  results in  Figure~\ref{fig:rb}. For both Armonk and Almaden (and for IBM's devices in general), the calibrations described above are performed every 24 hours. Our experiments ran around-the-clock via a cloud job queuing system, with varying elapsed time to the prior calibration.
\section{Compiler Flow} \label{sec:compiler_flow}

\begin{table}
\renewcommand{\arraystretch}{1.3}
\begin{tabular}{p{0.14\textwidth}p{0.193\textwidth}l@{}}
\toprule
Stage & Notes & Example \\
\midrule
Programming Language & High-level; hardware-unaware; sophisticated control flow & \texttt{qft(qc)} \\
Assembly & Usually 1- or 2- qubit arity gates; minimal control flow  & \texttt{h q[0]} \\
Basis Gates & Like an HDL; hardware-aware gate set & \texttt{u\textsubscript{1}(3) q[0]} \\
Pulse Schedule & Analog waves across channels; ultimate ``at the metal'' control & \raisebox{-0.65\totalheight}{\includegraphics[height=2.5\fontcharht\font`\B]{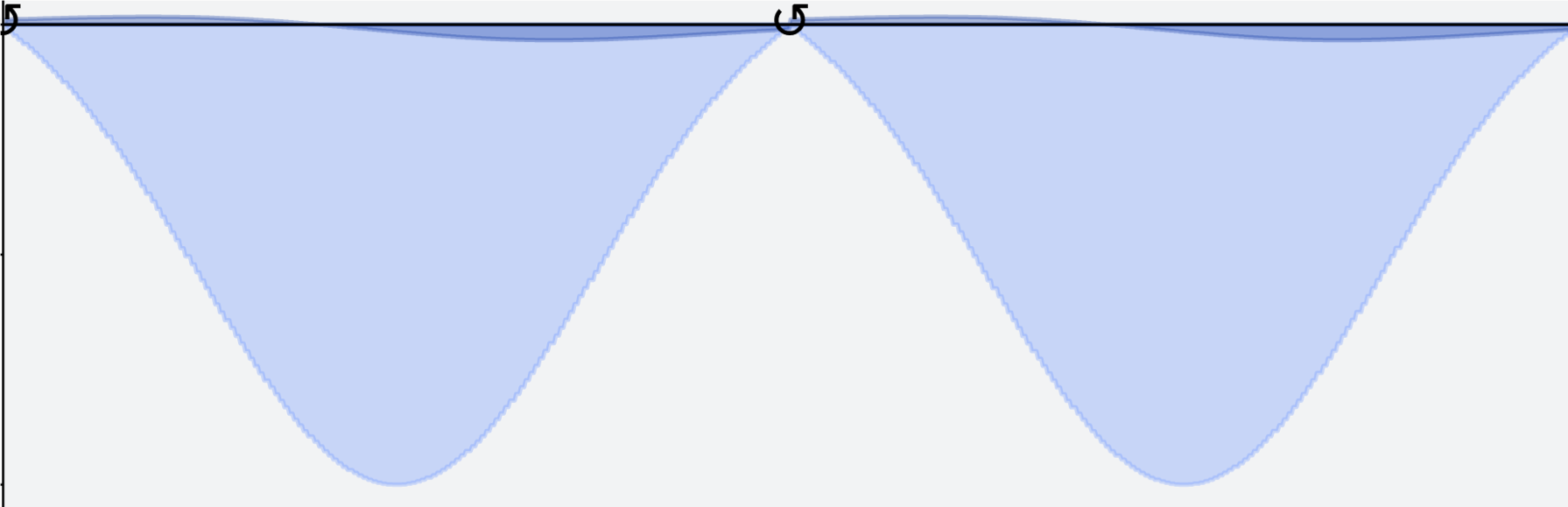}}
 \\
\bottomrule
\end{tabular}
\caption{Summary of the four stages of a quantum compiler.} \label{tab:compiler_stages}
\end{table}

As depicted in Figure~\ref{fig:compilation_flow}, quantum compilation proceeds through four stages, from high-level to low-level: programming language, assembly, basis gates, and pulse schedule.  Also shown is our alternative flow, which creates an augmented set of basis gates and a more optimized pulse schedule.  Table~\ref{tab:compiler_stages} presents a summary of these four stages. We now discuss existing implementations of these stages, why we should augment the standard set of basis gates, our new compiler framework, and the tradeoffs we considered when we designed the framework.

\subsection{Standard Flow}
\subsubsection{Programming Language}
Quantum PLs are designed to be user-friendly, with sophisticated control flow, debugging tools, and strong abstraction barriers between target operations and underlying quantum hardware. The most successful languages have been implemented as Python packages, such as IBM's Qiskit \cite{Qiskit}, Google's Cirq \cite{Cirq}, and Rigetti's PyQuil \cite{smith2016practical}. Others are written as entirely new languages, such as Scaffold \cite{abhari2012scaffold, javadiabhari2014scaffcc} which is based on LLVM infrastructure; Quipper \cite{green2013quipper} which is a functional language embedded in Haskell; and Q\# \cite{svore2018q} which is Microsoft's quantum domain specific language.

\subsubsection{Assembly} Quantum assembly languages are closer to hardware, but still aim to be device-agnostic. Generally, the assembly instructions only allow 1- or 2- qubit arity, since hardware primitives act on only 1 or 2 qubits at a time\footnote{The notable exception is trapped ion quantum computers, which support global entangling operations that simultaneously act on $N$ qubits \cite{sorensen2000entanglement, maslov2018use}.}. Quantum assembly is essentially equivalent to the quantum circuit representation of quantum programs. Prominent examples include OpenQASM \cite{mckay2018qiskit}, Rigetti's pyQuil \cite{smith2016practical}, and TUDelft's cQASM \cite{khammassi2018cqasm}.

\subsubsection{Basis Gates}
Basis gates are similar to assembly, but re-expressed in terms of the gate set that hardware implements. For example, while the well-known Controlled-$Z$ instruction is valid in assembly code, it would be re-written in basis gates as a sequence of $H$ and $CNOT$ gates---which hardware natively implements. The distinction between assembly and basis gates is primarily a conceptual one; in Qiskit, Cirq, and PyQuil, the basis gate and assembly layers are expressed in the same software framework. In some other domains, for example the Blackbird language \cite{killoran2019strawberry} for continuous-variable quantum computing, the assembly already resembles a hardware-aware basis gate layer. Regardless of the relationship between between assembly and basis gates, our core observation in this paper is that existing implementations of basis gate sets are too far from pulse-level hardware primitives. We will expand on this observation for the rest of the paper.

\subsubsection{Pulse Schedule}
\label{subsec:pulse_schedule}
The ultimate lowest-level control of a quantum computer is a schedule of complex-valued analog pulses, across multiple input channels. The image in Table~\ref{tab:compiler_stages} shows a sample pulse schedule on a single channel. The input channels are controlled by an Arbitrary Waveform Generator (AWG) which outputs a continuous value on each channel at every dt. Modern AWGs, such as the one in our experimental realization using IBM's Almaden system, achieve 4.5 Gigasamples per second, i.e. a new complex number every 0.22 ns.

The pulse schedule on drive channel $j$ is referred to as $d_j(t)$ and is complex-norm constrained by $|d_j(t)| \leq 1$. However, qubits are not directly acted on by $d_j(t)$ or $\operatorname{Re}[d_j(t)]$. Instead, the $d_j(t)$ signal is mixed with a \textit{local oscillator} of frequency $f_j$, leading to a final signal
\begin{equation}
D_j(t) = \operatorname{Re}[d_j(t) e^{i f_j t}] \label{eq:awg}
\end{equation}
This equation will be relevant when we demonstrate qudit operations in Section~\ref{sec:qudit_operations}.

The translations from basis gates to pulse schedules are known analytically. For example, in superconducting quantum hardware, $R_z$ basis gates are implemented in software with zero-duration and perfect-accuracy via the virtual-Z-Gate translation \cite{mckay2017efficient, krantz2019quantum}. The $X$ basis gates is transformed into almost-Gaussian ``DRAG'' pulses \cite{motzoi2009simple, gambetta2011analytic, motzoi2013improving}. In the OpenPulse interface, these translations are stored in the \texttt{cmd\_def} object, and reported by the hardware.

\begin{scriptsize}
\renewcommand{\arraystretch}{2.0}

\tabulinesep=1.2mm
\begin{table*}
  \centering
  \caption{Costs of various two qubit operations, by Native gate. Cost reductions at the right indicate optimization opportunities. One $\sqrt{\text{iSWAP}}$ is treated as 0.5 cost, while iSWAP has 1.0 cost.} \label{tab:operation_native_costs}
  \begin{tabu}{|p{20mm}|p{24mm}|l|llll|l|l|}
    \multicolumn{2}{l}{} & \multicolumn{7}{c}{Decomposition Cost by Native Gate} \\ \cline{3-9} 
    \multicolumn{2}{l|}{} & ``Textbook'' & \multicolumn{4}{c|}{Discrete Gates} & Half & \textbf{Parametrized} \\ \cline{1-2} 
    \textbf{Operation} & Standard Circuit Rep. & CNOT & CR($90^{\circ}$) & iSWAP & bSWAP & MAP & $\sqrt{\text{iSWAP}}$ & \textbf{CR(}$\bm{\theta}$\textbf{)} \\
    \hline \hline
    CNOT & \multicolumn{1}{|c|}{\adjustbox{valign=t}{\Qcircuit @C=1em @R=0.4em { & \ctrl{1} & \qw \\ & \targ & \qw}}} & 1 & 1 & 2 & 2 & 1 & 1 & \textbf{1} \\ \hline
    SWAP & \multicolumn{1}{|c|}{\adjustbox{valign=t}{\Qcircuit @C=1em @R=0.4em { & \ctrl{1} & \targ & \ctrl{1} & \qw \\ & \targ & \ctrl{-1} & \targ & \qw}}} & 3 & 3 & 3 & 3 & 3 & 1.5 & \textbf{3} \\ \hline
    ZZ Interaction & \multicolumn{1}{|c|}{\adjustbox{valign=t}{\Qcircuit @C=1em @R=0.4em { & \ctrl{1} & \qw & \ctrl{1} & \qw \\ & \targ & \gate{R_z} & \targ & \qw}}} & 2 & 2 & 2 & 2 & 2 & 1 & \textbf{1} \\ \hline
    Fermionic \quad Simulation & \multicolumn{1}{|c|}{\adjustbox{valign=t}{\Qcircuit @C=0.3em @R=0.2em { & \gate{R_z} & \multigate{1}{\text{iSWAP}} & \ctrl{1} & \gate{R_z} & \qw \\ & \gate{R_z} & \ghost{\text{iSWAP}} & \gate{Z} & \gate{R_z} & \qw}}} & 3 & 3 & 3 & 3 & 3 & 1.5 & \textbf{3} \\ \hline
  \end{tabu}
\end{table*}
\end{scriptsize}

\subsection{Motivation for Different Basis Gates} \label{subsec:motivation}
At a high level, our core observation is that existing basis gates sets are too far from actual hardware primitives at the pulse-level. This leads to missed opportunities for optimization. Sections~\ref{sec:direct_drive}--\ref{sec:qudit_operations} will present optimizations resulting from specific gaps between basis gates and pulse-level hardware primitives. Table~\ref{tab:operation_native_costs} introduces one such gap that we expand upon in Section~\ref{sec:cross_resonance_optimizations}. Each row in the table is a two-qubit operation. The columns express the cost\footnote{Cost here means the number of two-qubit gates needed, since they dominate both error and duration.} of performing the target operation using the given native gate. We computed these costs using Qiskit's \texttt{TwoQubitBasisDecomposer} tool, which uses the KAK decomposition \cite{khaneja2001cartan} described further in \cite{cross2019validating}.

The CNOT column indicates the number of CNOT gates needed to implement the target operation. CNOT is the default ``textbook'' two-qubit gate, so algorithms are usually written in terms of CNOT. The next group of four columns, Discrete Gates, captures basis gates from
\begin{itemize}
    \item Fixed-frequency superconducting qubits: $90^{\circ}$ Cross-Resonance \cite{paraoanu2006microwave, rigetti2010fully, chow2011simple}, bSWAP \cite{poletto2012entanglement}, and MAP \cite{chow2013microwave}.
    \item Frequency-tunable superconducting qubits: iSWAP \cite{chow2010quantum} and also CZ \cite{strauch2003quantum, dicarlo2009demonstration} which is omitted because it is equivalent to CNOT.
    \item Quantum dot spin qubits: iSWAP \cite{loss1998quantum}
    \item Nuclear spin qubits: iSWAP \cite{mozyrsky2001indirect}
\end{itemize}
All four of these columns have identical costs to the CNOT column. As a result of this parity, the prevailing sentiment in current quantum compilation software is that these basis gates are equivalent. Moreover, since quantum algorithms are usually written in terms of CNOTs, there is not an obvious reason to deviate from these basis gates.

The two rightmost columns, challenge this sentiment. The $\sqrt{\text{iSWAP}}$ reflects the fact that quantum hardware allows one to perform ``half'' of an iSWAP by damping the pulse shape of a standard iSWAP gate. This Half-gate leads to significant improvements over full iSWAPs--each row's cost is halved. The CNOT decomposition and SWAP decomposition are known \cite{echternach2001universal, lebedev2018extended}, but to the best of our knowledge, the ZZ Interaction (ubiquitous operation for quantum chemistry and optimization algorithms)  and Fermionic Simulation (ubiquitous for quantum chemistry) decompositions are not previously known. They will have immediate applications on hardware that supports $\sqrt{\text{iSWAP}}$ such as frequency-tunable superconducting qubits, quantum dot spin qubits, and nuclear spin qubits.

The rightmost column, bolded because it was our experimental target, reflects the fact that fixed-frequency superconducting qubits support parametrized Cross-Resonance($\theta$) via pulse stretching. Since the native gate is parametrized, we used a different approach to compute the decomposition costs in its column. Specifically, we used the COBYLA constrained optimizer \cite{powell2007view} in Scipy \cite{scipy}, with the constraint of finding a 99.9+\% fidelity decomposition. Subject to this constraint, our decomposer minimizes the cost of the CR($\theta$) gates needed to perform the target operation. Observe that ZZ Interaction is 2x cheaper with a Parametrized CR($\theta$) gate than with the standard CR($90^{\circ}$) gate. The ZZ Interaction is in fact the most common two qubit operation in near-term algorithms. This optimization is expanded upon in Section~\ref{sec:cross_resonance_optimizations}.

Our method is extensible to other systems including trapped ions. Some of the trapped ion decompositions have already been studied in recent publications \cite{hempel2018quantum, muller2011simulating, nam2019ground}.

\subsection{Design of Our Compiler}
\label{subsec:our_compiler}
Our compiler is implemented as a fork of Qiskit. While Qiskit has traditionally been used in conjunction with IBM superconducting quantum computers, it is a generic framework that supports any underlying quantum hardware. For example, trapped ion quantum computer vendors have recently integrated with Qiskit \cite{javadiqiskit2019}, and OpenPulse support was recently added \cite{nguyen2020enabling} to the XACC infrastructure for quantum-classical computing \cite{mccaskey2020xacc}. Thus our framework is general, though we performed our experimental realizations on IBM hardware, which is the first to implement OpenPulse.

Our compiler maintains the overall structure of Qiskit, which is already designed with extensibility in mind. As discussed previously, we augment the set of basis gates to better match pulse-level primitives. To support this augmented basis gate set, we re-write the decomposition rules from assembly instructions to basis gates and add new translations (to \texttt{cmd\_def}) that convert augmented basis gates to pulse schedules. We expand on the augmented basis gates in Sections~\ref{sec:direct_drive}--\ref{sec:qudit_operations}.

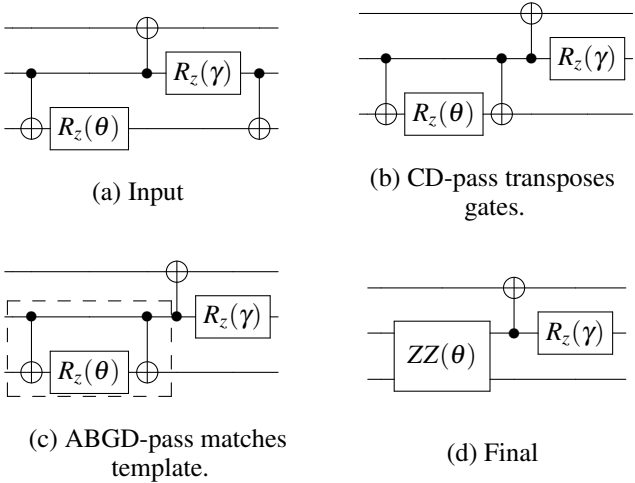
\begin{figure}
    
\begin{minipage}{0.2\textwidth}
\Qcircuit @C=0.3em @R=0.5em {
&\qw  &\qw        &\qw                  &\targ      &\qw & \qw &\qw\\
&\qw  &\ctrl{1}   &\qw                  &\ctrl{-1}  &\gate{R_z(\gamma)}               & \ctrl{1} &\qw\\
&\qw  &\targ      &\gate{R_z(\theta)}   &\qw        &\qw              & \targ &\qw
} 
\vspace*{0.4cm}
\begin{center}
    (a) Input
\end{center} 
\end{minipage}
\hfill
\begin{minipage}{0.2\textwidth}
\Qcircuit @C=0.3em @R=0.5em {
&\qw  &\qw        &\qw              & \qw    &\targ      &\qw  &\qw\\
&\qw  &\ctrl{1}   &\qw & \ctrl{1} &\ctrl{-1}  &\gate{R_z(\gamma)}        &\qw\\
&\qw  &\targ      &\gate{R_z(\theta)}  & \targ  &\qw        &\qw              &\qw
} 
\vspace*{0.4cm}
\begin{center}
    (b) CD-pass transposes gates.
\end{center} 
\end{minipage}
\vspace*{0.5cm}

\begin{minipage}{0.23\textwidth}
\Qcircuit @C=0.3em @R=0.5em {
&\qw  &\qw        &\qw              & \qw    &\targ      &\qw  &\qw\\
&\qw  &\ctrl{1}   &\qw & \ctrl{1} &\ctrl{-1}  &\gate{R_z(\gamma)}        &\qw\\
&\qw  &\targ      &\gate{R_z(\theta)}  & \targ  &\qw        &\qw              &\qw
\gategroup{2}{3}{3}{5}{1em}{--}} 
\vspace*{0.4cm}
\begin{center}
    (c) ABGD-pass matches template.
\end{center} 
\end{minipage}
\hfill
\begin{minipage}{0.22\textwidth}
\hspace*{0.2cm}
\Qcircuit @C=0.5em @R=0.5em {
&\qw  &\qw     &\targ      &\qw  &\qw\\
&\qw  &\multigate{1}{ZZ(\theta)} &\ctrl{-1}  &\gate{R_z(\gamma)}        &\qw\\
&\qw  &\ghost{ZZ(\theta)}  &\qw        &\qw              &\qw
} 
\vspace*{0.4cm}
\begin{center}
    (d) Final
\end{center} 
\end{minipage}
    \caption{Depiction of our compiler passes for commutativity detection (CD) and augmented basis gate detection (ABGD).}
    \label{fig:transpiler_passes}
\end{figure}

To take advantage of the augmented basis gates, we added Qiskit transpiler passes, which convert input quantum assembly into optimized quantum assembly in the spirit of LLVM Transform passes. Our transpiler passes \textbf{automatically} optimize user code by using the augmented basis gates. One transpiler pass traverses a DAG-representation of the quantum assembly and pattern matches for templates that represent sequences of gates (such as the ZZ Interaction) that reduce to an augmented basis gate. We also include a commutativity detection transpiler pass that performs this pattern matching even when obfuscated by false dependencies in intermediate gates; this pass is inspired by techniques described in \cite{shi2019optimized}. Figure~\ref{fig:transpiler_passes} shows an example of these two passes. Through these two passes, we maintain the ``write-once target-all'' behavior of user-written code, which can remain hardware agnostic.

\subsection{Compiler Design Tradeoffs}

Another compiler design we considered is Quantum Optimal Control \cite{werschnik2007quantum, Chow2010ImplementingOC}, which translates directly from the programming language (specifically from the quantum circuit's overall unitary matrix) down to highly optimized pulses. QOC has been explored extensively in physics communities and more recently from an architectural perspective \cite{shi2019optimized, gokhale2019partial}.

QOC is indeed a promising path for future machines, and in fact our original aim was to perform pulse-shaping via optimal control. However, our experience revealed experimental roadblocks. In particular, QOC requires a perfect characterization of the quantum computer's underlying physics, i.e. the device Hamiltonian. Pulses designed from an inaccurate Hamiltonian accumulate substantial error. Moreover, to be experimentally realistic, QOC-generated pulses must be constrained to have bounded amplitudes and smooth derivatives. These constraints diminish both the potential advantage of QOC and the reliable convergence of QOC algorithms \cite{leung2017speedup, gokhale2019partial}. In addition, optimizing the pulse shape requires evaluation of partial derivatives of a fidelity metric---a task that is easy analytically or in simulation, but extremely difficult with noisy experimental measurements.

Our experience is mirrored by other work on QOC---the vast majority of prior work has been performed via simulation. The few experimental realizations of QOC generally focus on state preparation (easier than unitary synthesis), e.g. \cite{song2019generation, omran2019generation}. Moreover, these experiments impose significant Hamiltonian tomography or calibration overhead, for example staggered field calibration \cite{omran2019generation}. Experimental realizations on superconducting qubits, whose Hamiltonians drift over time \cite{murali2019noise, tannu2019not}, are even more rare. In fact, the state-of-art for pulse shaping on superconducting qubits has eschewed standard QOC entirely \cite{leng2019robust}, focusing instead on a closed-loop feedback for tuning pulses. We refer to \cite{leng2019robust} for further details on the experimental barriers (and opportunities) to QOC, particularly in superconducting qubits. We also note that recent progress in robust control \cite{ball2020software, qctrl_singlequbitgates, qctrl_crossresonance} is promising and could justify QOC-based approaches in future work. The work in \cite{ball2020software} is already compatible with OpenPulse.

Our approach to pulse-shaping arose from these limitations. In particular, our techniques are bootstrapped from the standard basis gate calibrations, which are already performed daily. By decomposing and then re-scaling the pre-calibrated pulses, we generate an augmented basis gate set, without ever requiring the device Hamiltonian. We emphasize that our technique can be applied on current cloud-accessible quantum devices, as documented in our Github repository \cite{anonymized_repo}. Moreover, while QOC generally leads to convoluted pulses, our pulses are very simple. This simplicity minimizes the possibility of control errors and also leads to greater interpretability.

\section{Optimization 1: Direct Rotations} \label{sec:direct_drive}
We now present the first of our four optimizations enabled by pulse control. The gist of this optimization is that pulse-level control enables us to perform single-qubit gates (qubit state rotations on the Bloch sphere) via a \textit{direct} trajectory, saving time and potentially reducing errors.

It can be shown that any arbitrary single-qubit gate, termed $U_3$ in Qiskit, can be implemented by tuning up a single pulse that rotates the qubit state by 90 degrees around the $X$ axis (the $R_x(90^{\circ})$ pulse). This is doable due to the following identity, and due to the fact that rotations about the $Z$ axis can be implemented in software at no cost (implemented by a compiler transformation on all future gates involving the target qubit)~\cite{mckay2017efficient}. 

\begin{equation}
\normalsize
U_3(\theta,\phi,\lambda) = R_z(\phi+90^{\circ}) R_x(90^{\circ}) R_z(\theta+180^{\circ}) R_x(90^{\circ}) R_z(\lambda)
\label{eq:u3_qiskit}
\end{equation}

The above is extremely attractive from a hardware calibration perspective, since it suggests that fine tuning one pulse is enough to achieve high-fidelity single-qubit gates. In fact, this is how these gates are implemented on IBM quantum computers. We now present experimental evidence that access to one more calibrated gate, as well as pulse control, gives the compiler the ability to optimize single-qubit gates further.

\subsection{Direct $X$ gates}
We first consider the simple $X$ operation, which acts as a NOT by flipping $\ket{0}$ and $\ket{1}$ quantum states. $X$ gates are ubiquitous in algorithms. Our approach relies on access to the $X=R_x(180^{\circ})$ rotation, which is already pre-calibrated, as discussed in Section~\ref{subsec:calibration}. In our experiments we had access to such a pulse, but one could also be calibrated by the user through OpenPulse. We emphasize that this extra pulse is not strictly necessary for universal computation. However, we use it to demonstrate the power of an \textit{overcomplete} basis for optimizations.

Qiskit's standard compilation flow decomposes an $X$ operation into a $U_3$ instruction per equation~\ref{eq:u3_qiskit}. At the pulse level, the $U_3$ instruction is implemented by two consecutive $R_x(90^{\circ})$ pulses. Together these complete an $X$ gate (i.e. $180^{\circ}$ rotation).

However, the indirection of implementing $X$ with two $R_x(90^{\circ})$ pulses becomes unnecessary in the presence of a pre-calibrated $R_x(180^{\circ})$ gate. The procedure for calibrating such a gate is very similar to the $R_x(90^{\circ})$, and its direct calibration has benefits beyond our discussion here~\cite{majer2007coupling, galperin2005rabi, ashhab2006rabi,sheldon2016procedure}.
On IBM hardware enabled with OpenPulse, this pulse is readily available in the backend pulse library.

In our compiler, we exploit this simple observation by augmenting the basis gates with a \texttt{DirectX} gate, which is linked to the $R_x(180^{\circ})$ pulse that is already calibrated on the quantum computer. This gate is twice as fast as Qiskit's standard $X$ gate, and has 2x lower error, as measured through quantum state tomography experiments.

\begin{figure}
     \centering
     \begin{subfigure}[b]{0.45\textwidth}
         \centering
         \includegraphics[width=\textwidth]{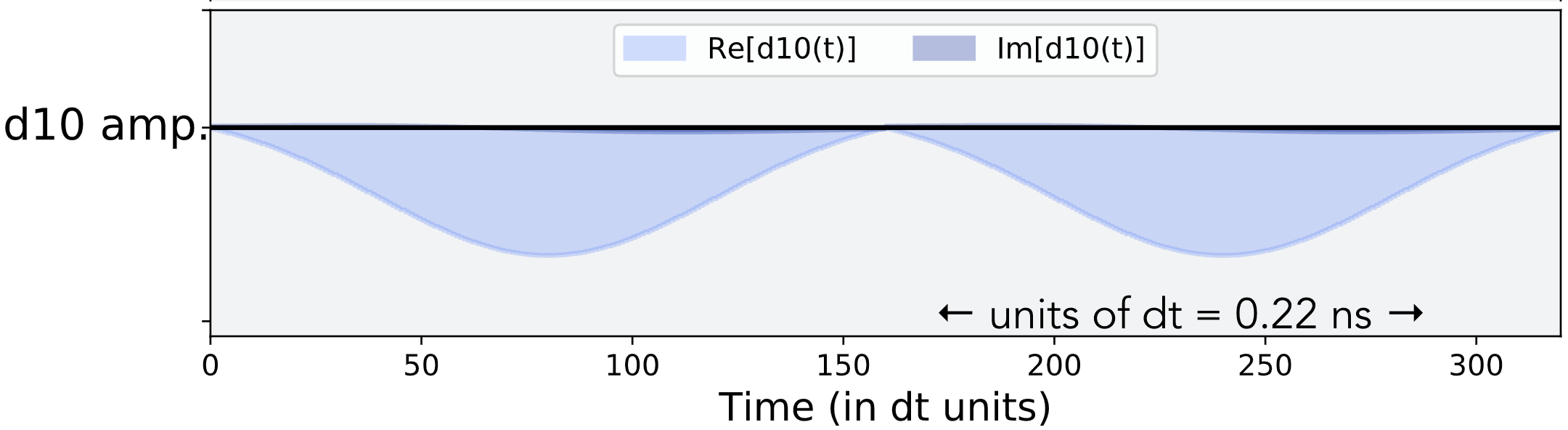}
     \end{subfigure}

\par\bigskip

\par\medskip

     \begin{subfigure}[b]{0.45\textwidth}
         \centering
         \includegraphics[width=\textwidth]{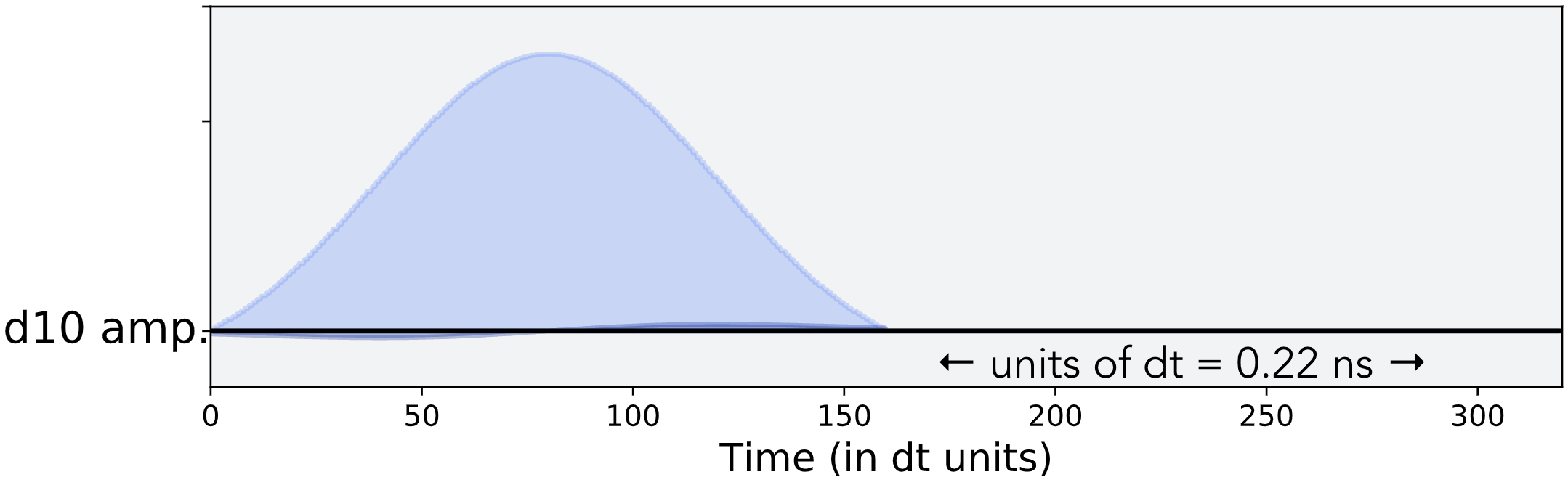}
     \end{subfigure}
    \caption{Pulse schedules for the $X$ gate via standard compilation (top) versus via direct compilation via our approach (bottom). Time is in units of dt = 0.22 ns. Thus, the \texttt{DirectX} gate takes 35.6 ns, twice as fast as the 71.1 ns standard $X$ gate.}
    \label{fig:standard_vs_direct_x}
\end{figure}

Figure~\ref{fig:standard_vs_direct_x} depicts a comparison of pulse schedules used to achieve the $X$ gate in 71.1 ns in the standard framework vs. 35.6 ns in our optimization.
It also illustrates why these two pulse schedules are logically equivalent: they have the same (absolute) area-under-curve. To a first approximation---which we will refine below---this area determines how much rotation is applied.

We next consider more sophisticated direct rotation gates, for general angles.

\subsection{Direct partial rotation about the $X$ axis}
Since OpenPulse gives us access to arbitrary pulse envelopes, it is natural to ask whether ``partial" rotations about the $X$ axis ($R_x(\theta)$ gates) can be realized more efficiently without invoking two discrete $R_x(90^{\circ})$ pulses (as done by the standard Qiskit decomposition in Equation~\ref{eq:u3_qiskit}). Our compiler does this by downscaling the amplitude of the pre-calibrated $R_x(180^{\circ})$ pulse by $\frac{\theta}{180^{\circ}}$ to achieve the $R_x(\theta)$ rotation. We represent this as the \texttt{DirectRx($\theta$)} augmented basis gate in our compiler. Since we rely on the pre-calibrated $R_x(180^{\circ})$ this technique imposes no calibration overhead.

\begin{figure}
     \centering
     \begin{subfigure}[b]{0.48\textwidth}
         \centering
         \includegraphics[width=\textwidth]{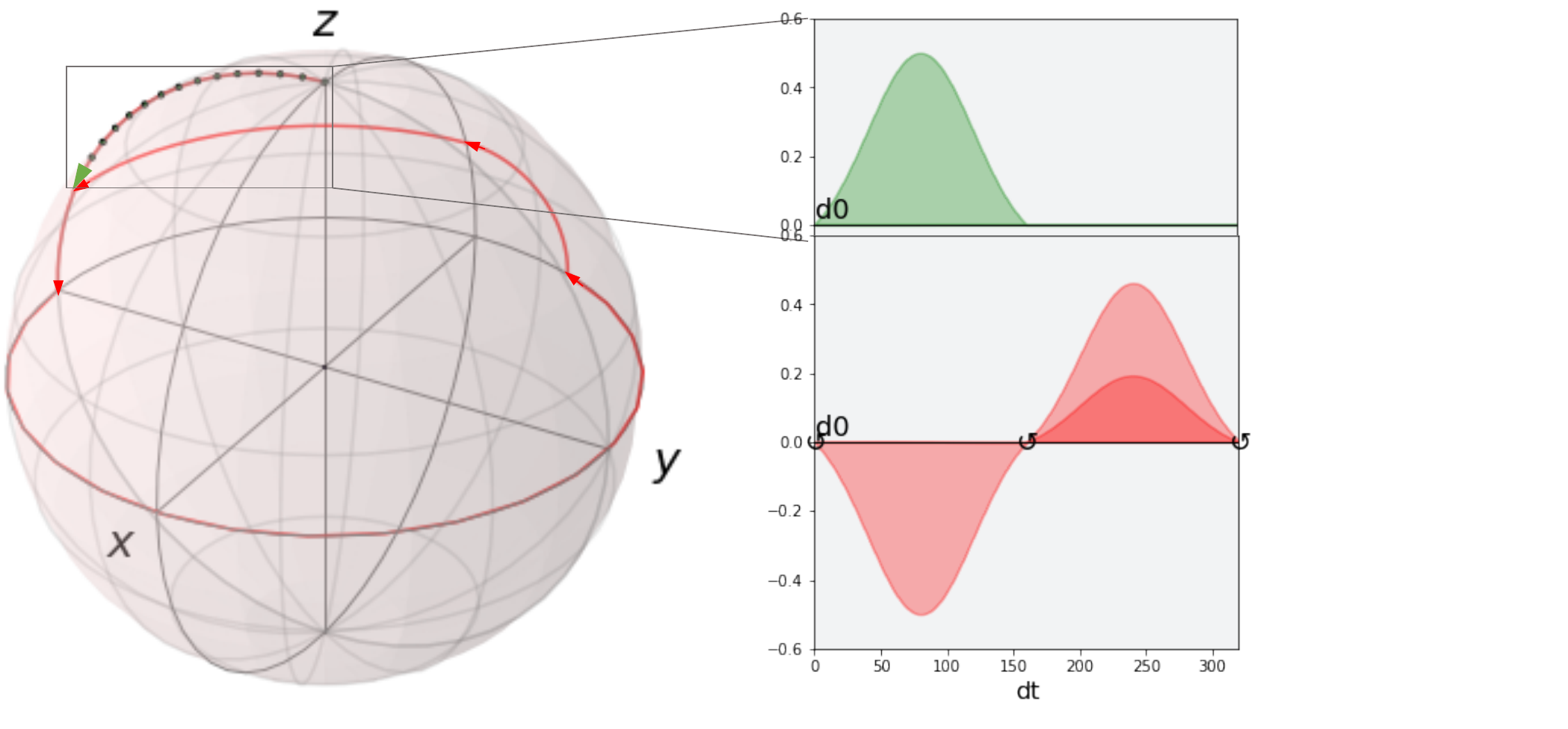}
    \label{fig:rx_sample}
     \end{subfigure}

\par\medskip

     \begin{subfigure}[b]{0.48\textwidth}
         \centering
         \includegraphics[width=\textwidth]{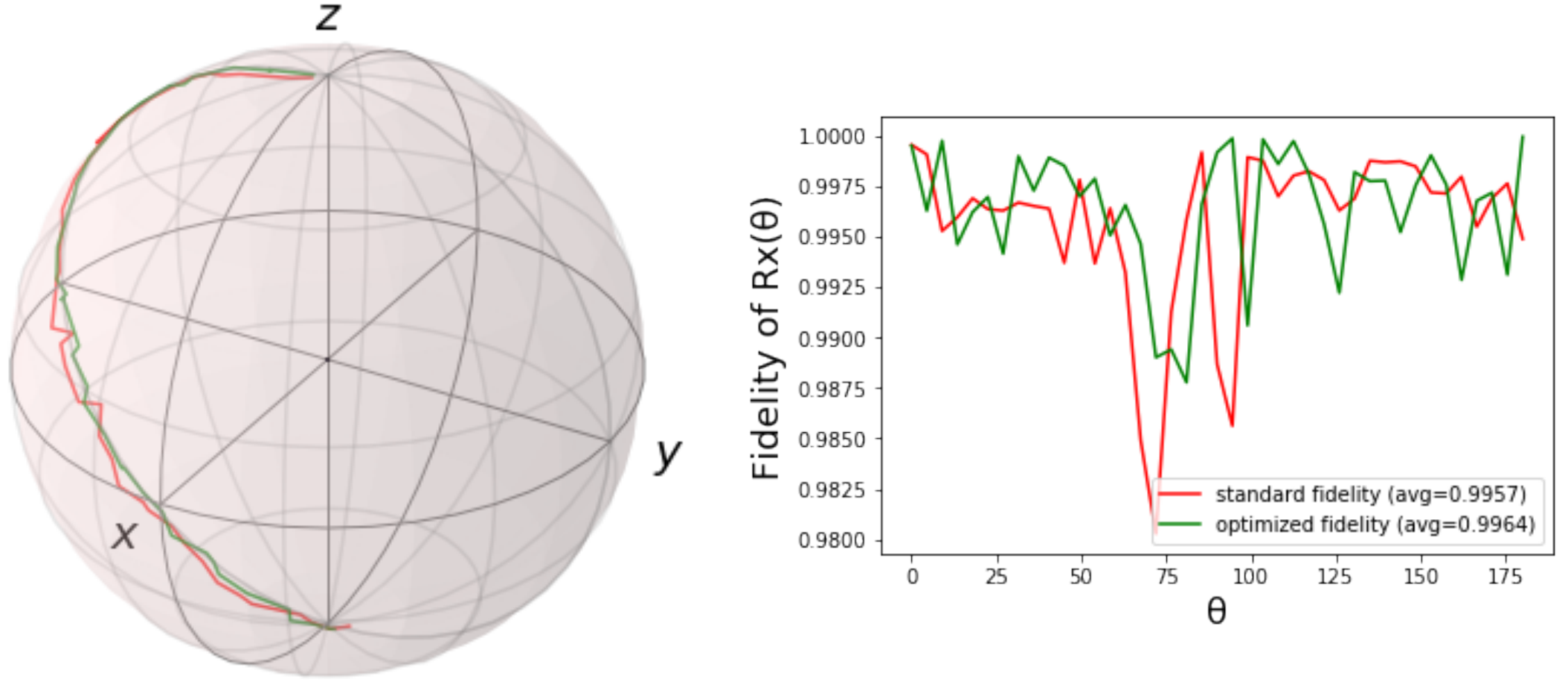}
     \label{fig:rx_all}
     \end{subfigure}
    \caption{Illustration of gate-level vs. pulse-level rotation about the X axis. (top) Trajectory of an $R_x(67^{\circ})$ rotation, and the pulses that implement them. Standard gate-based compilation (red) includes two applications of the pre-calibrated $R_x(90^{\circ})$ pulse (interleaved with $R_z$ (frame changes) which are zero-cost and in software). Optimized pulse-based compilation takes the shortest path from origin to destination, with only one scaled pulse.
    (bottom) Fidelity of $R_x(\theta)$ rotations. Each data point is obtained using quantum state tomography experiments to rotate around the $X$ axis by $\theta$. Standard gate-compiled rotations (red) show more jitter from ideal, and 16\% higher error on average, compared to optimized pulse-compiled rotations (green).}
    \label{fig:rx}
\end{figure}

The results of our experiments with the new \texttt{DirectRx($\theta$)} are summarized in Figure~\ref{fig:rx}. Bypassing the gate abstraction, our technique speeds up all $R_x$ rotations by 2x and has 16\% lower error on average. We discuss the source of the error reduction in Section~\ref{subsec:improvement_source}.

In the next subsection, we will note how \texttt{DirectRx($\theta$)} generalizes to arbitrary-axis rotations for free.

\subsection{Optimizing generic rotations}
Equipped with an augmented gate set that implements arbitrary $X$ axis rotations at reduced cost, we now show that all single-qubit gates can be achieved with one pulse. Recall that in standard Qiskit compilation, general single qubit gates are implemented via two $R_x(90^{\circ})$ pulses and three no-cost $R_z$ frame changes. However, we can write the same gate as~\cite{nielsen2002quantum}:

\begin{equation}
U_3(\theta,\phi,\lambda) = R_z(\phi+180^{\circ}) R_x(\theta) R_z(\lambda-180^{\circ})
\end{equation}

Recall that $R_z$ rotations are implemented by frame changes 
%(i.e. phase shifts between real and complex parts of input waveforms) 
with perfect fidelity and 0 duration.  Thus, this implies that any single-qubit gate can be performed using direct $R_x(\theta)$ rotations, sandwiched by free $R_z$ gates.

\subsection{Compiler implications}
In the preceding subsection, we showed how an augmented gate set can be beneficial. However, the compiler now has more than the minimum set of pulses to work with to realize a quantum gate. In order to decide which pulses to use when, we need a deeper understanding and characterization of the errors incurred by $R_x(\theta)$ gates for arbitrary $\theta$. We can use this system characterization to inform the compiler about the best pulse substitution strategy.

We performed pulse simulations and real experiments to gain insight into the errors. Our simulations were done using Qiskit's OpenPulse simulator. We enhanced the simulator to find the Hamiltonian terms for IBM's Almaden system, through a reverse-engineering process and fitting the results to the device-reported pulse library.

Taking Almaden's pre-calibrated direct $X$ pulse (DRAG pulse), we scaled the area-under-curve down by a factor of $\frac{0}{40}, \frac{1}{40}, \frac{2}{40}, ..., 1$. To first order, these should perform $R_x(\theta)$ for $\theta = 0^{\circ}, 4.5^{\circ}, 9^{\circ}, ..., 180^{\circ}$. For each angle, we performed three simulations and three experiments to measure the $X$, $Y$, and $Z$ components of the final quantum state, which allows us to plot on the Bloch sphere.

\begin{figure}
     \centering
     \begin{subfigure}[b]{0.26\textwidth}
         \centering
         \includegraphics[width=\textwidth]{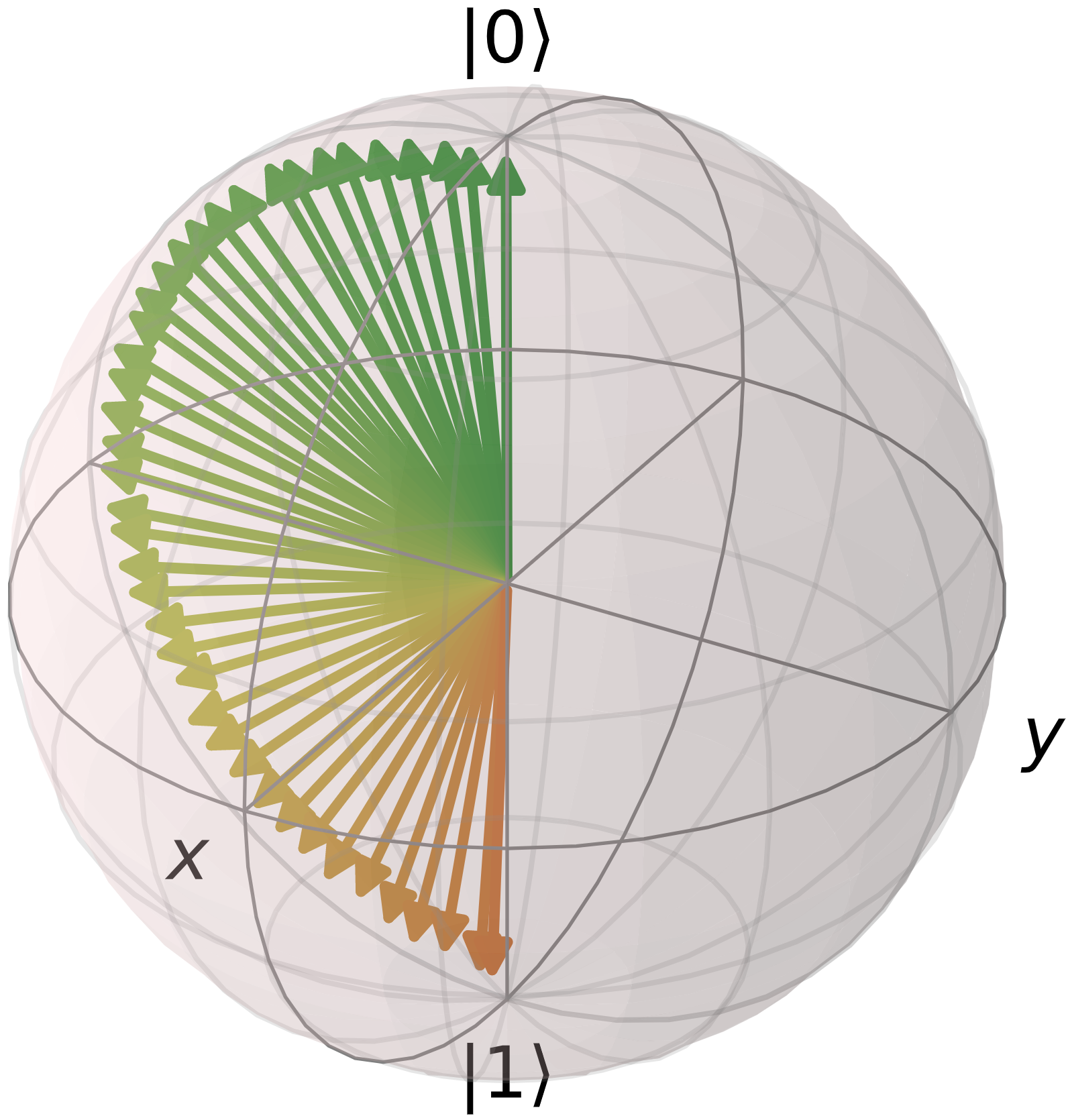}
         \caption{Sweeping 41 angles from $\theta=0^{\circ}$ in green to $\theta=180^{\circ}$ in orange.}
        \label{fig:direct_rx_simulated_bloch}
     \end{subfigure}
     \hfill
     \begin{subfigure}[b]{0.177\textwidth}
         \centering
         \includegraphics[width=\textwidth]{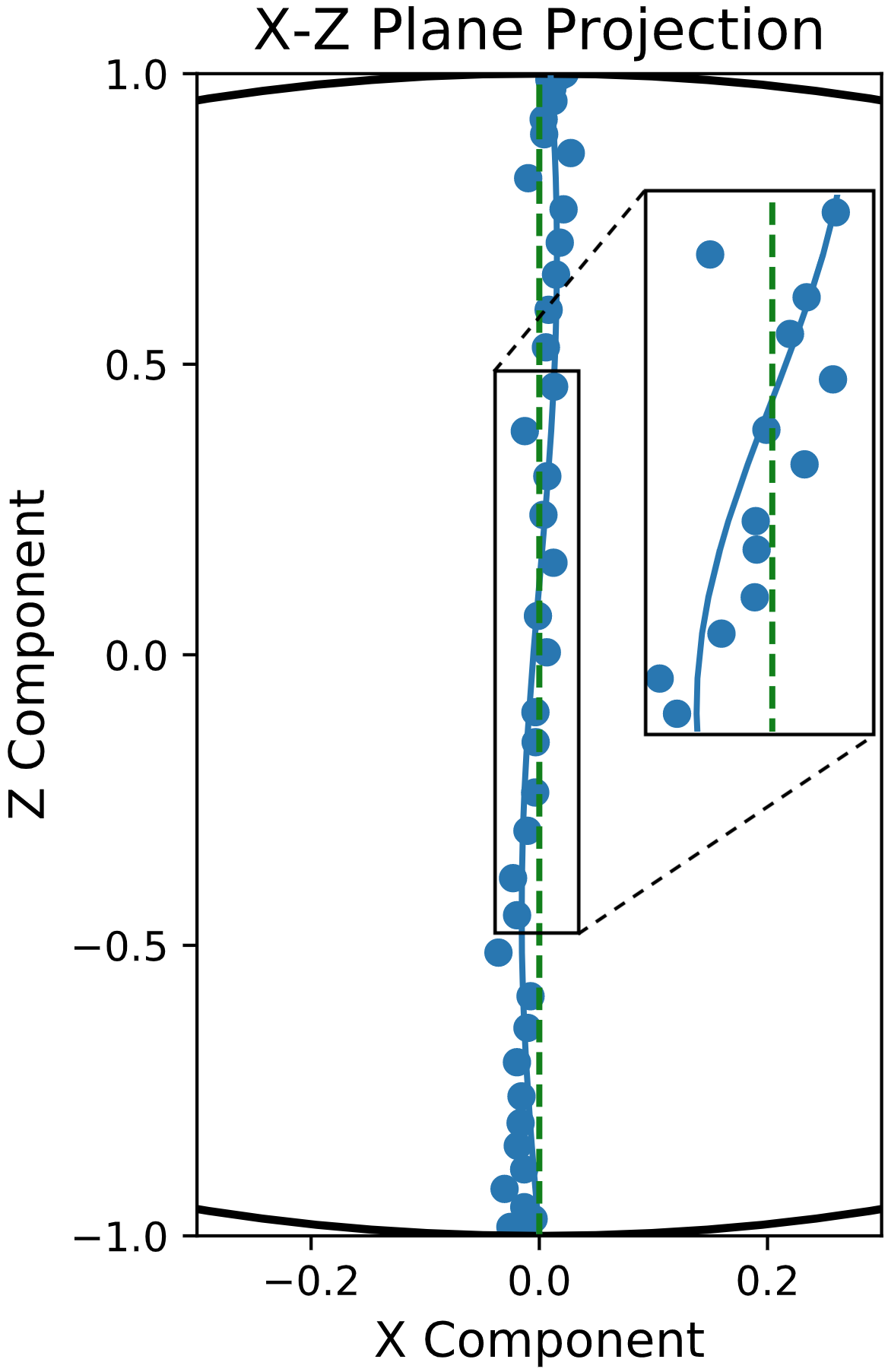}
         \caption{XZ trajectory slightly deviates from $X = 0$.}
        \label{fig:direct_rx_simulated_xz}
     \end{subfigure}
        \caption{Simulated results for Direct $R_x(\theta)$. The inset magnifies the $X$ component.} \label{fig:direct_rx_simulated}
\end{figure}

Figure~\ref{fig:direct_rx_simulated} depicts the results of simulation. Plotting only the X-Z plane, we see that deviations from the Prime Meridian are quite small, but do have a sinusoidal pattern (at exactly $0^{\circ}$, $90^{\circ}$, and $180^{\circ}$, there is no dephasing). These simulation results are in agreement with an independent simulation from~\cite{krantz2019quantum}.

\begin{figure}
     \centering
     \begin{subfigure}[b]{0.26\textwidth}
         \centering
         \includegraphics[width=\textwidth]{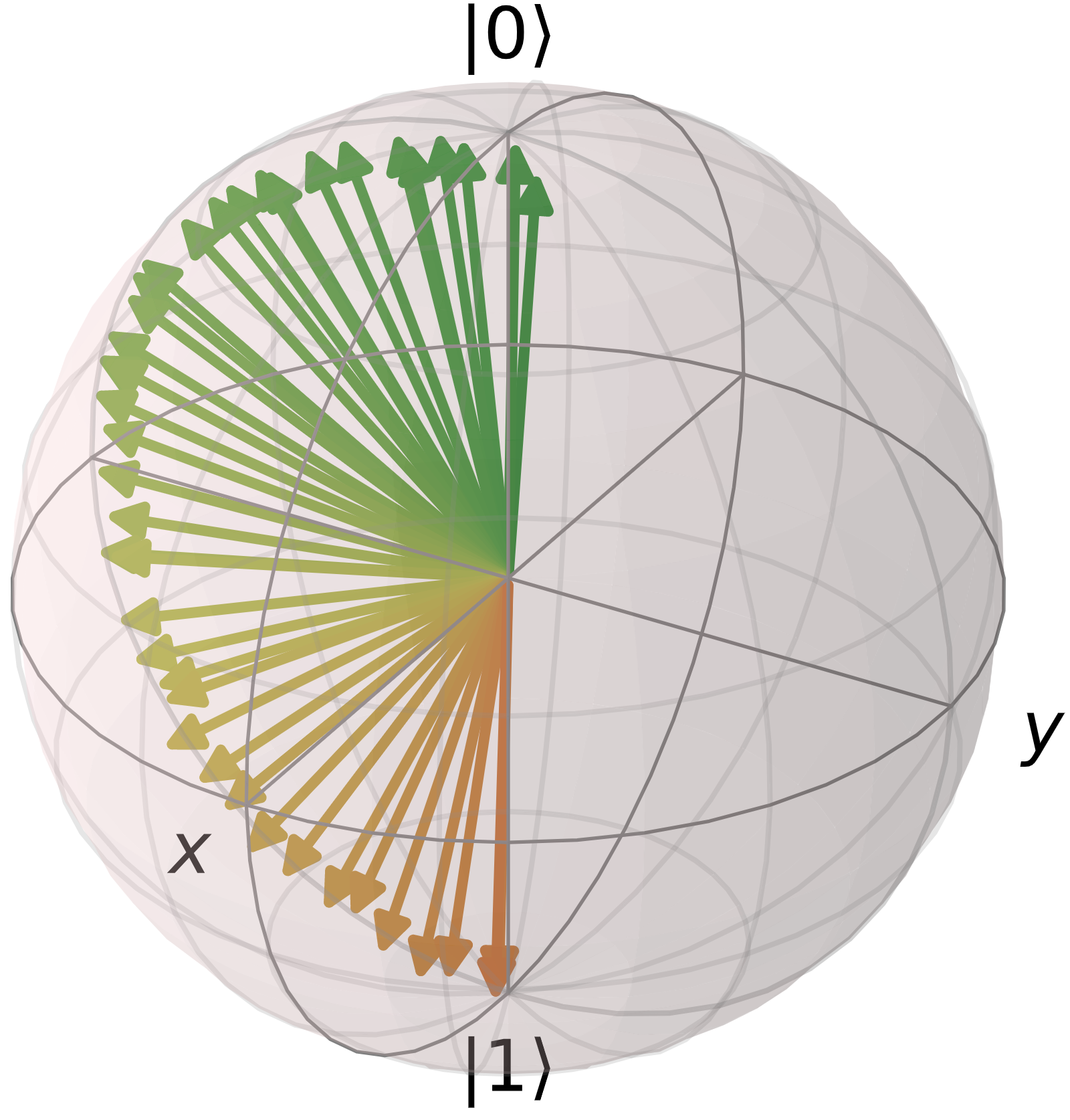}
     \end{subfigure}
     \hfill
     \begin{subfigure}[b]{0.177\textwidth}
         \centering
         \includegraphics[width=\textwidth]{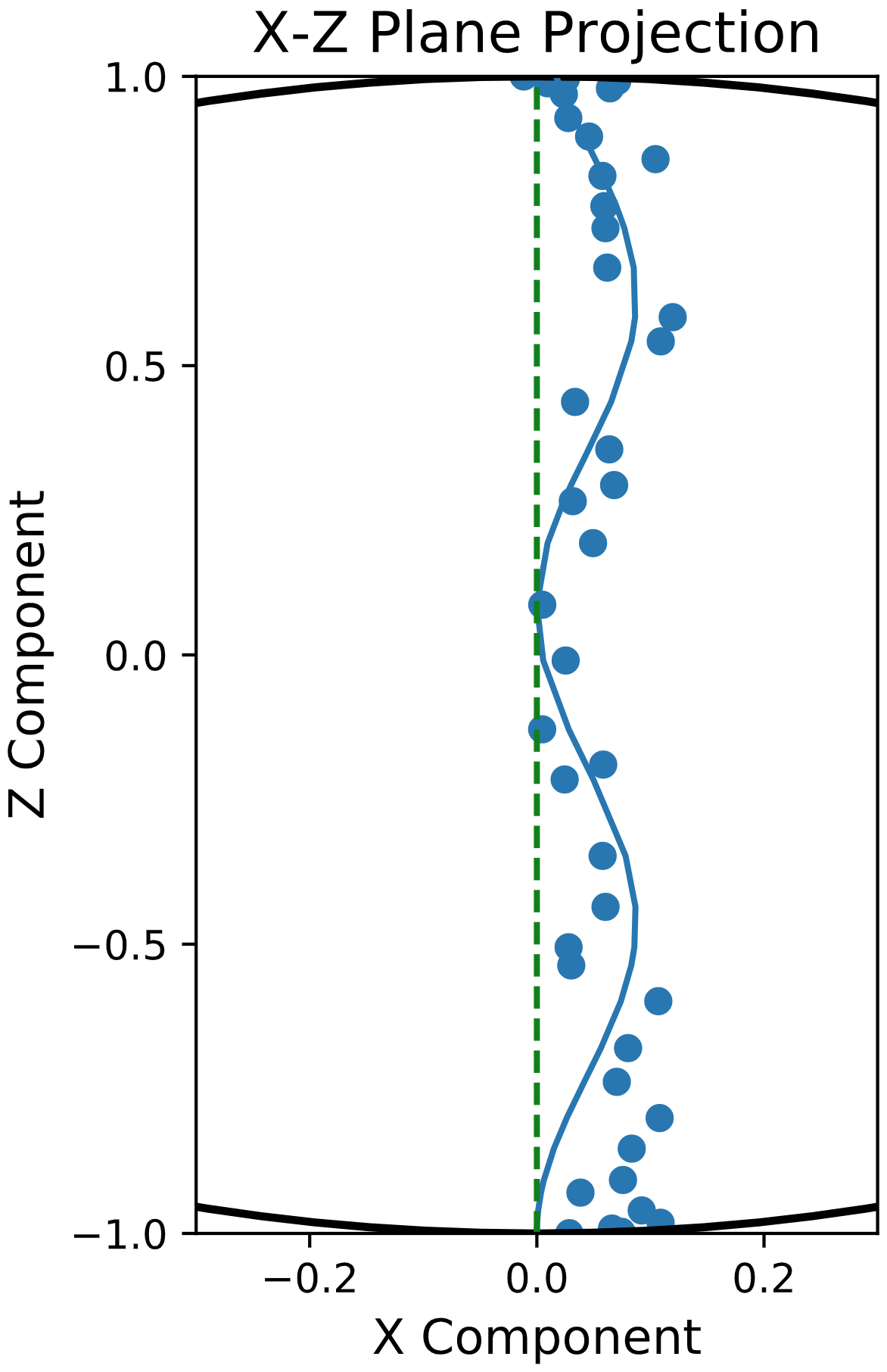}
     \end{subfigure}
        \caption{Experimental results for Direct $R_x(\theta)$ on IBM's Almaden system, based on $3 \times 41 \times 1000 = 123$k shots. This empirical characterization of dephasing from the Meridian can be used to make the gate better at each $\theta$.}
        \label{fig:direct_rx_experiment}
\end{figure}

The experimental results are presented in Figure~\ref{fig:direct_rx_experiment}. We note two deviations from simulation: (1) the $X$ components are still sinusoidal but now translated to the right and (2) the magnitude of the $X$-component deviations are larger. However, we can treat these characterization results with an empirical attitude---now that we know the dephasing at each $\theta$ value point, we can perform an $R_x(\theta)$ gate by applying a scaled-down $X$ pulse, and then correcting the phase error in accordance with the data in Figure~\ref{fig:direct_rx_experiment}.

\section{Optimization 2: Cross-Gate Pulse Cancellation} \label{sec:cross_pulse_cancellation}
The gist of this optimization is that standard basis gates are not atomic\footnote{We use atomic in the common-usage sense of something that cannot be decomposed into something else more fundamental. This should not be confused with technical meanings of atomicity in computing.}, despite conveying this perception. By augmenting basis gates with the true atomic primitives, new gate cancellation opportunities emerge that lead to 24\% speedups for common operations.

\subsection{Theory}
Generally, two-qubit basis gates are not atomic. For example, in Qiskit, the CNOT basis gate is implemented at the pulse level as a combination of single qubit gates, plus invocations of the hardware primitive Cross-Resonance pulse:

$$
\Qcircuit @C=0.5em @R=2em {
& \ctrl{1} & \qw \\
& \targ & \qw
}
\quad = \quad
\Qcircuit @C=1em @R=1em {
& \gate{X} & \multigate{1}{CR(-45^{\circ})} & \gate{X} & \multigate{1}{CR(45^{\circ})} & \qw \\
& \gate{R_x(90^{\circ})} & \ghost{CR(-45^{\circ})} & \qw & \ghost{CR(45^{\circ})} & \qw
} \qquad
$$

Notice in particular, that even the invocation of the hardware primitive Cross-Resonance pulse is not a clean atomic unit, but is decomposed into two pulses separated by an $X$ gate. This ``echoed'' Cross-Resonance pulse design is necessary to perform a $CR(90^{\circ})$ gate (which is the generator of CNOT) with high fidelity \cite{corcoles2013process}.

This analysis reveals there are opportunities for gate cancellation on either side of the CNOT\footnote{The circuit decomposition clearly depicts gate cancellation opportunities on the left side of the CNOT with the $X$ and $R_x(90^\circ)$ gates; alternatively, the top $X$ can be shifted rightward by commutation identities to create cancellation opportunities on the right side}. In fact, such sequences are common. To enable these cancellations, we augment the basis gate set with the hardware primitive $CR(\pm 45^{\circ})$ basis gates, which are free from pre-calibrated CNOTs. We replace the assembly instruction for CNOT into this decomposition and invoke Qiskit's optimizer to perform gate cancellations.

\begin{figure}
     \centering
     \begin{subfigure}[b]{0.37\textwidth}
         \centering
         \includegraphics[width=\textwidth]{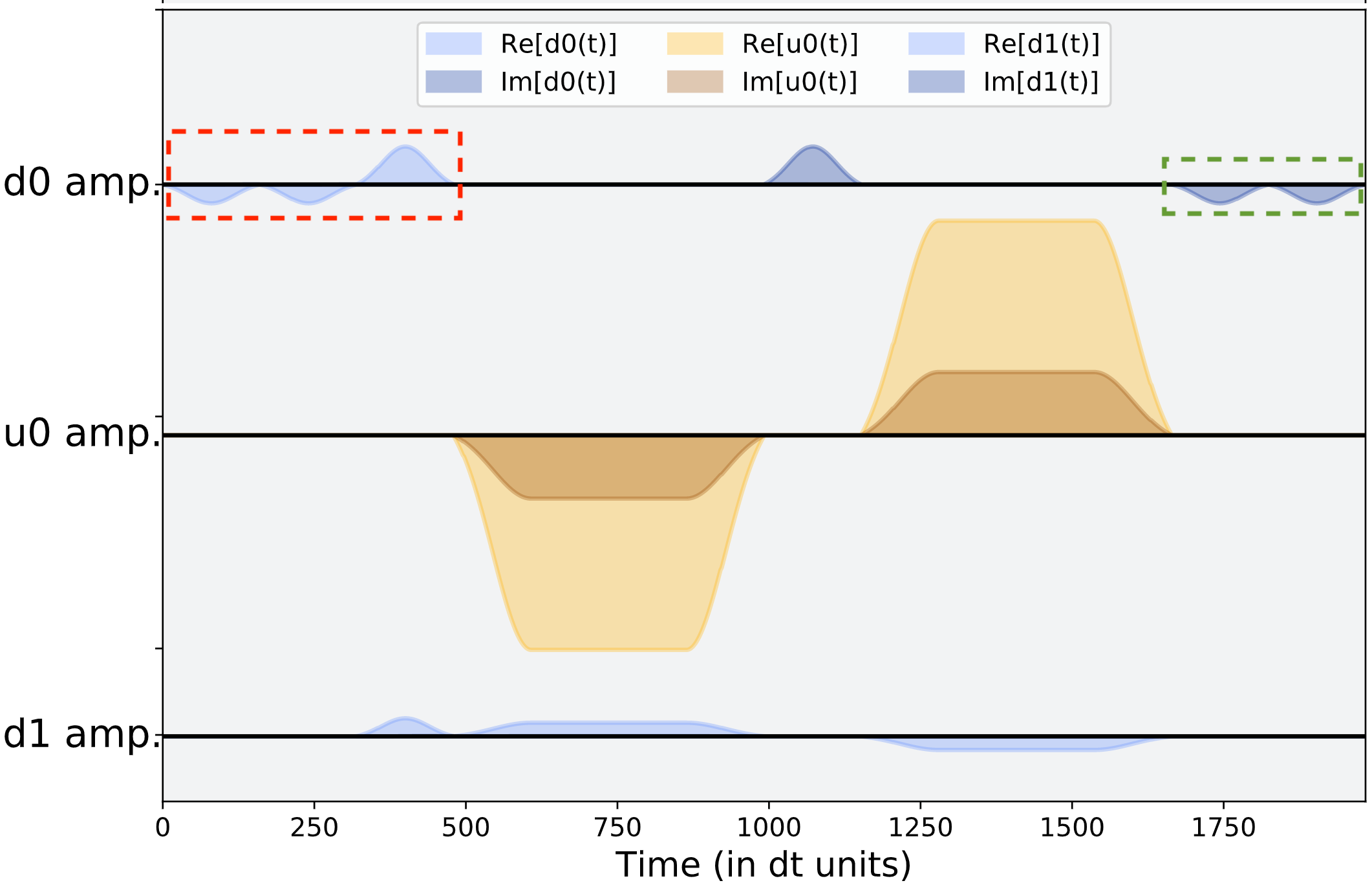}
     \end{subfigure}

\par\bigskip

\par\medskip

     \begin{subfigure}[b]{0.39\textwidth}
         \centering
         \includegraphics[width=\textwidth]{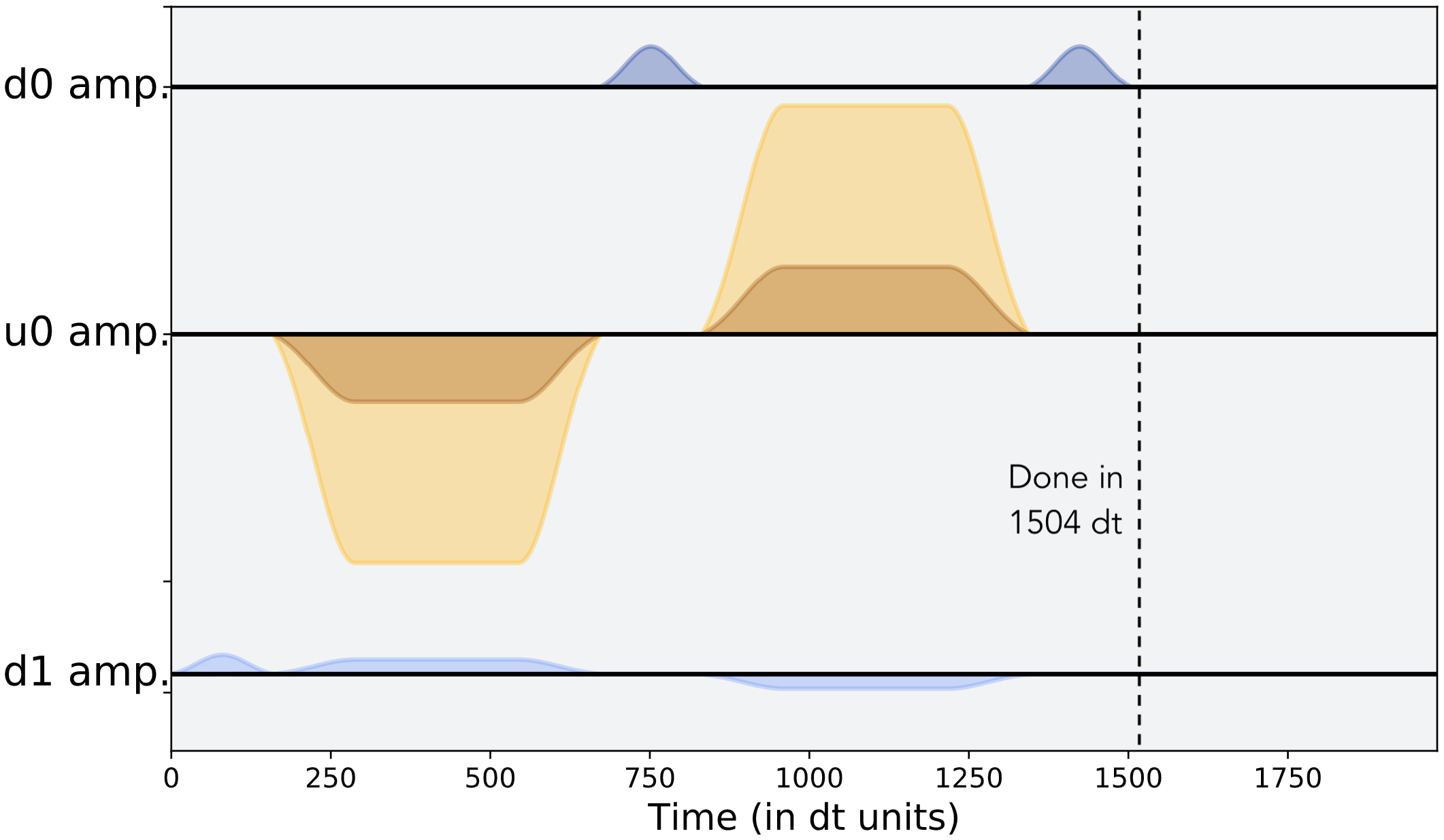}
     \end{subfigure}
    \caption{Pulse schedules for the open-CNOT by standard compilation (top) and our optimized compilation (bottom). Our compiler cancels out the $X$ rotation gates in the red box and combines the two $R_x(-90^{\circ})$ pulses in the green box into a single $R_x(180^{\circ}) = X$ pulse. This reduces the total duration by 24\% from 1984 dt to 1504 dt.}
    \label{fig:open_cnot_schedules}
\end{figure}

\subsection{Application}
To demonstrate our technique, we benchmarked using a common operation: the open-Controlled-NOT. The open-CNOT has the ``opposite'' behavior as a CNOT: it flips the target if the control is $\ket{0}$ and does nothing if the control is $\ket{1}$. Its implementation via the CNOT basis gate is simple: first an $X$ on the control, then a standard CNOT, and then another $X$ to restore the control.

However, by decomposing the CNOT into our augmented basis gates, the first $X$ on the control cancels with the ``internal'' $X$ in the decomposition of CNOT. Figure~\ref{fig:open_cnot_schedules} depicts the pulse schedules for the open CNOT under standard compilation (top) and via our compilation (bottom). Notice that two $X$'s in the red box cancel out, leading to a 24\% reduction in runtime.

We tested the open-CNOT pulse schedules experimentally. To isolate the effect of cross-gate pulse cancellation, we performed the direct $X$ gate from the previous section in both variants. The resulting data indicates a modest increase in success probability from 87.1(9)\% to 87.3(9)\%, measured over 16k shots (hence the Bernoulli standard deviation of 0.09\%). We emphasize that the open-CNOT is just one of many typical quantum operations that have $R_x$ rotations next to two-qubit basis gates. Our compiler takes advantage of all such cancellation opportunities, which are otherwise invisible at the granularity of standard, non-atomic basis gates.
\section{Optimization 3: Two Qubit Optimizations} \label{sec:cross_resonance_optimizations}
The gist of this optimization is that standard basis gates lead to inefficient decompositions of important two-qubit operations. Instead, we can use pulse-level hardware primitives as new basis gates that lead to operations with 60\% lower error.

\subsection{Theory}
Recall from Table~\ref{tab:operation_native_costs} that two-qubit operations can be achieved by using a ``half'' or parametrized basis gate set. For example, data movement (SWAP) is 2x more costly on superconducting qubits with an iSWAP basis gate than on qubits with a $\sqrt{\text{iSWAP}}$ basis gate.

Here, we study basis gate decompositions using the parametrized Cross-Resonance pulse $CR(\theta)$, which is the pulse-level hardware primitive on IBM devices. However, we again emphasize that our compiler techniques immediately generalize to any other basis gate decompositions.

As discussed in Section~\ref{sec:cross_pulse_cancellation}, neither $CR(\theta)$ nor even $CR(90^{\circ})$ are exposed as standard basis gates. Our compiler first extracts the pulse for the $CR(90^{\circ})$ gate from the \texttt{cmd\_def} pulse schedule for the CNOT basis gate. Then, to implement $CR(\theta)$ for arbitrary $\theta$, we horizontally stretch the $CR(90^{\circ})$, guided by knowledge of IBM's specific ``active cancellation echo'' implementation of the Cross-Resonance pulse \cite{sheldon2016procedure, magesan2018effective}.

\begin{figure}
    \begin{subfigure}[b]{0.48\textwidth}
        \includegraphics[width=\textwidth]{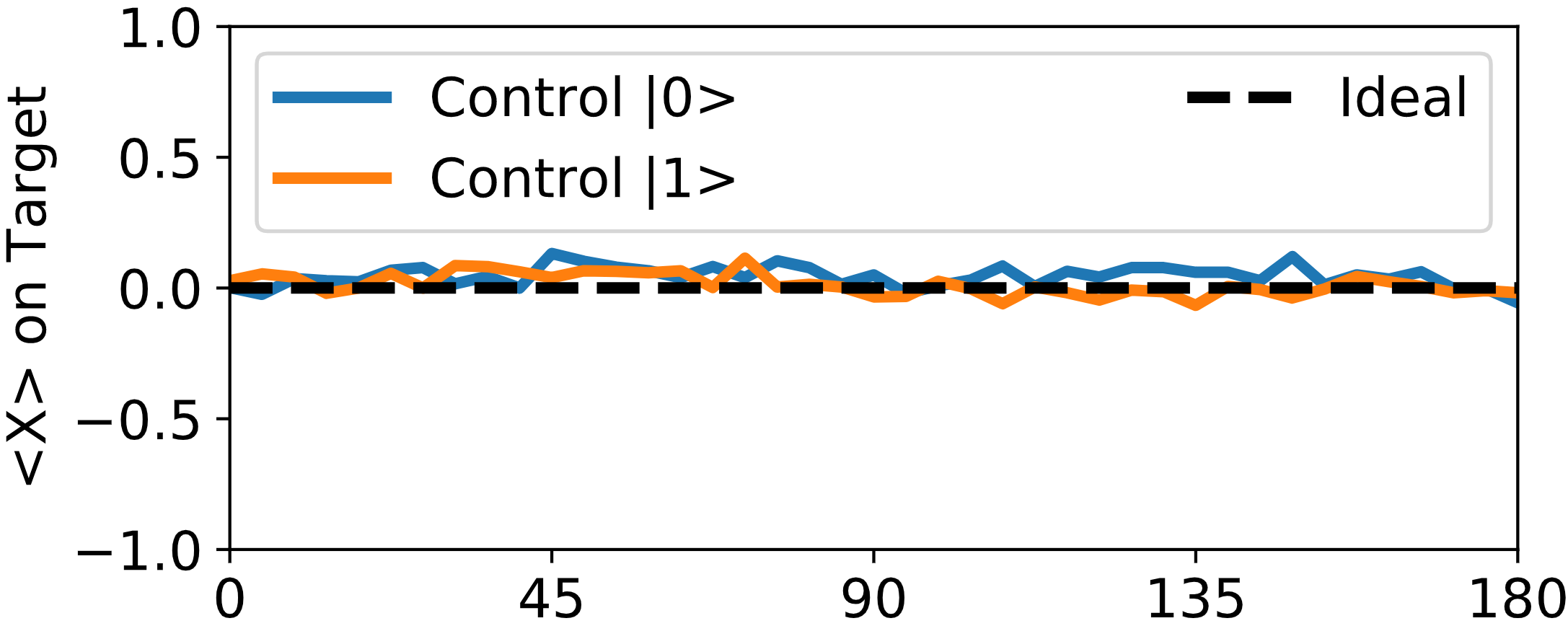}
    \end{subfigure}
    \begin{subfigure}[b]{0.48\textwidth}
        \includegraphics[width=\textwidth]{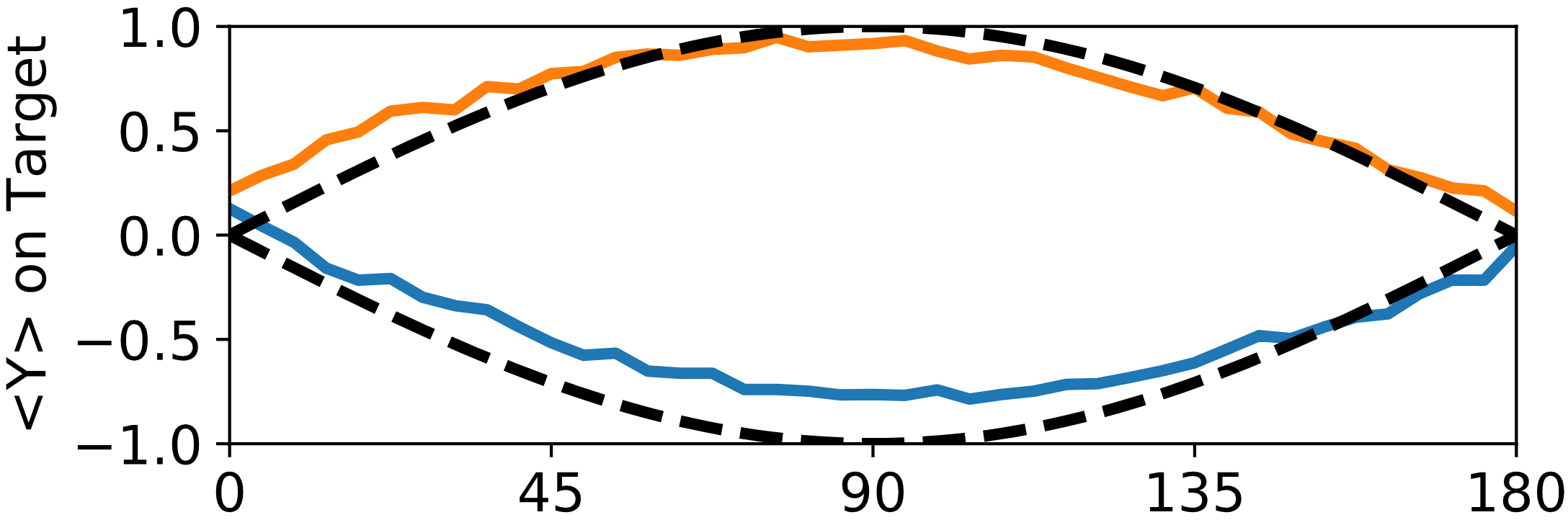}
    \end{subfigure}
    \begin{subfigure}[b]{0.48\textwidth}
        \includegraphics[width=\textwidth]{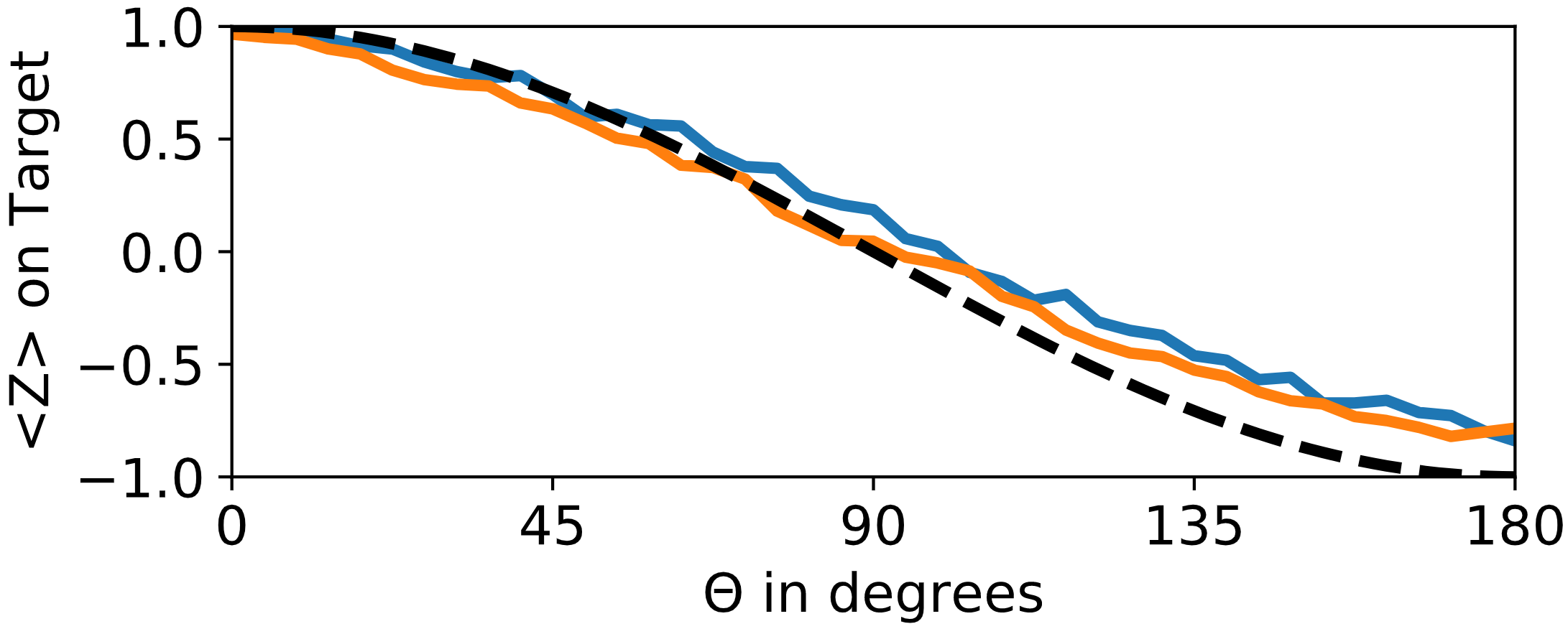}
    \end{subfigure}
    \caption{Tomography on the target qubit in Cross-Resonance($\theta$) pulse. Results from both experiment and simulation agree with ideal results. $41 \times 3 \times 2 \times 1000$ = 246k shots.}
 \label{fig:target_experiment}
\end{figure}

Figure~\ref{fig:target_experiment} shows our experimental results, which closely track with the ideal curve. Given the successful implementation of $CR(\theta)$ at the pulse level, we added it as a new basis gate.

\subsection{Application}

As indicated by the last row of Table~\ref{tab:operation_native_costs}, the ``ZZ Interaction'' two-qubit operation can be implemented using a single CR($\theta$) gate. By contrast, the ``textbook'' implementation using standard basis gates requires two CNOTs. The CR($\theta$) decomposition is depicted below. While this decomposition is fairly simple in hindsight, we discovered it computationally using the optimization procedure mentioned in Section~\ref{subsec:motivation}.
$$
\Qcircuit @C=1.3em @R=.8em {
& \ctrl{1} & \qw           & \ctrl{1} & \qw \\
& \targ    & \gate{R_z(\theta)} & \targ    & \qw
} \quad =
\quad \Qcircuit @C=1.3em @R=1.em {
& \qw & \multigate{1}{CR_{\theta}} & \qw & \qw \\
& \gate{H} & \ghost{CR_{\theta}}    & \gate{H}    & \qw
}
$$

\begin{figure}
    \centering
    \includegraphics[width=0.48\textwidth]{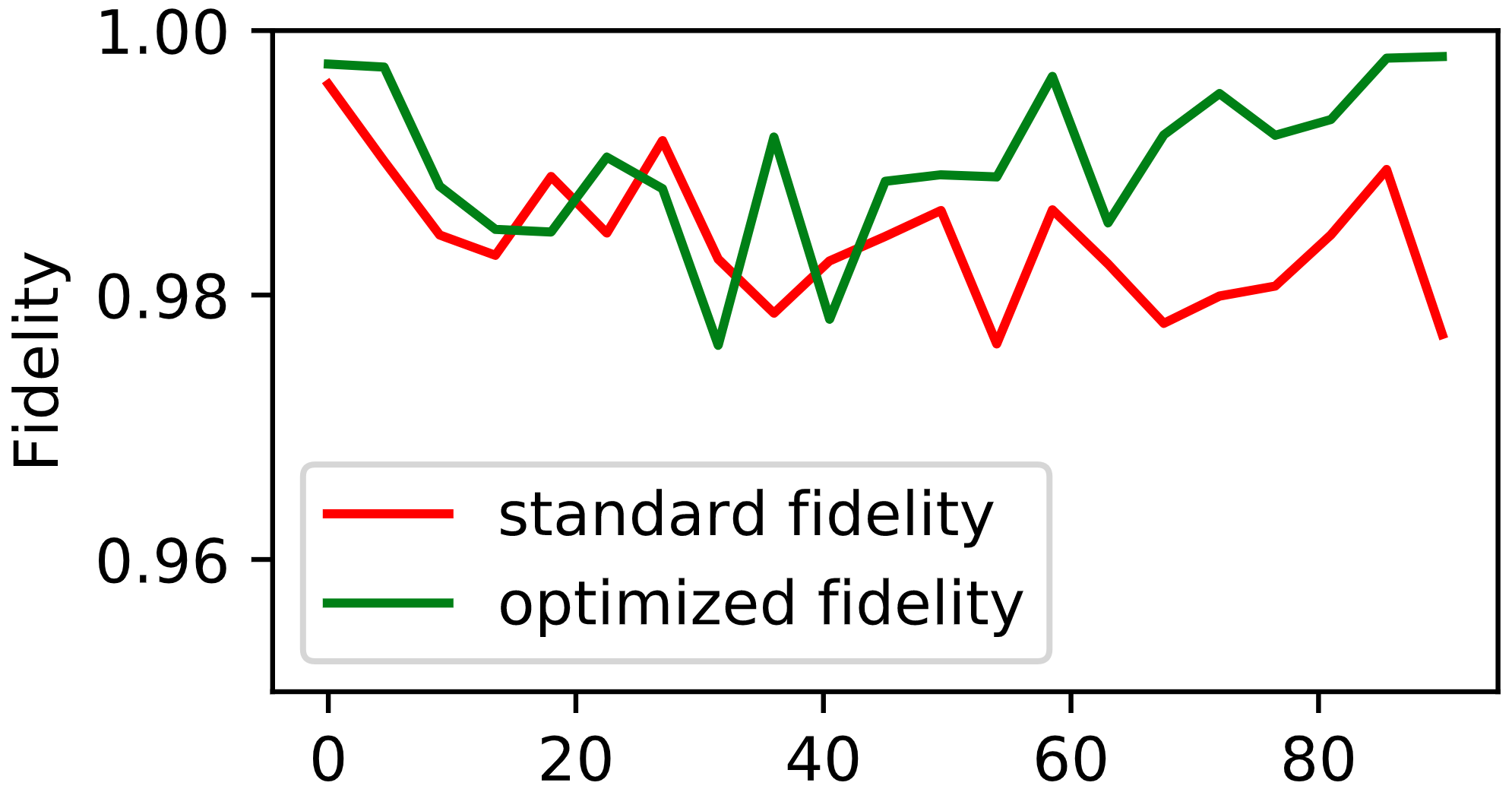}
    \caption{Experimental results for state fidelity, measured for the ZZ Interaction by standard compilation vs. our optimized compilation. These results reflect $21 \times 2 \times 2000$ = 84k shots. Standard and optimized have fidelities of 98.4\% and 99.0\% respectively. Thus, our compiler achieves an average 60\% reduction in error for the ZZ Interaction.}
    \label{fig:zz_fidelity}
\end{figure}

To experimentally verify our ZZ Interaction technique, we implemented it using both the standard compiler (i.e. CNOT, $R_z(\theta)$, CNOT) and our optimized compiler (H, CR($\theta$), H) for $\theta$ spanning from $0^{\circ}$ to $90^{\circ}$ in $4.5^{\circ}$ increments. As shown in Figure~\ref{fig:zz_fidelity}, our compiler achieves better results, with a 60\% average reduction in error (state infidelity).

As we will see in Section~\ref{sec:results}'s Benchmark Results, the ZZ Interaction is the most frequent two-qubit operation in near-term algorithms. Thus, this optimization is the dominant source of improvements in full benchmarks. Before continuing, we re-iterate that our compiler passes (as discussed in Section~\ref{subsec:our_compiler}) automatically identify ZZ Interactions in user-code, even when obfuscated by false data dependencies. Therefore, programmers may continue to write code using ``textbook'' CNOT decompositions and do not need to reason about device physics.

% This identity works, because
% $$
% \Qcircuit @C=1.3em @R=1.3em {
% & \multigate{1}{CR_{\theta}} & \qw  \\
% & \ghost{CR_{\theta}} & \qw
% } \qquad
% =
% \qquad \Qcircuit @C=1.3em @R=1.3em {
% & \ctrl{1} & \ctrlo{1} & \qw \\
% & \gate{R_x(-\theta)} & \gate{R_x(\theta)} & \qw
% } \qquad
% =
% \qquad \Qcircuit @C=1.3em @R=1.3em {
% & \qw & \ctrl{1} & \qw \\
% & \gate{R_x(\theta)} & \gate{R_x(- 2 \theta)} & \qw
% }
% $$

\section{Optimization 4: Qudit Operations} \label{sec:qudit_operations}
The gist of this optimization is that access to quantum hardware at the pulse level enables us to control energy states outside the qubit subspace. In particular, we can instead control our information carriers as d-level \textit{qudits}. We experimentally demonstrate this idea, by cycling a base-3 counter using a single \textit{qutrit}, a task that would be impossible with a single qubit. The counter achieves high fidelity, suggesting practical near-term applications.

\subsection{Theory}

Many quantum systems used to realize a qubit have other energy levels present, which can be used to construct quantum gates \cite{peterer2015coherence,ribeiro2019accelerated,earnest2018realization} or, as we demonstrate in this paper, to realize d-level \textit{qudits}. Substantial prior work observed an ``information'' compression advantage from using 3-level qutrits or higher level qudits \cite{pavlidis2017arithmetic}, which has been further applied to specific algorithms such as Grover search \cite{fan2008applications, li2011fast, wang2011improved, ivanov2012time} and Shor factoring \cite{bocharov2017factoring}. More recent work \cite{gokhale2019asymptotic} has even demonstrated exponential gains from using qutrits to implement common operations like the Generalized Toffoli.

However, across nearly all quantum hardware and associated software, standard basis gates are only written to address the qubit subspace of hardware. This is the case in part because the local oscillator described in Section~\ref{subsec:pulse_schedule} is set to oscillate at the energy gap between the $\ket{0}$ and $\ket{1}$ energy states, $f_{01}$. Since higher level states are separated by different energy gaps, under normal operation, gates can only address this qubit subspace.

However, we can circumvent this limitation by carefully designing our pulse schedule. For example, suppose we want to address the $\ket{1}$ to $\ket{2}$ transition subspace, whose energy gap we denote as $f_{12} = f_{01} + \alpha$. Per Equation~\ref{eq:awg}, applying a $d_j(t) = e^{- i \alpha t}$ pulse yields a total output of $e^{i f_{12} t}$. Thus, by designing a frequency-shifting pulse schedule, we can change the effective frequency of the local oscillator and target subspaces beyond the $\ket{0}$ to $\ket{1}$ regime.

\subsection{Application}

By transitioning to these higher energy levels one at a time we can realize a base-$d$ ``counter''. Not only is this a good benchmark for qudit control, it has potential application in both the near-term era of quantum computing and beyond. In the near-term, parity checks are commonplace \cite{mcardle2019error} (though most parity checks are for even/odd) and a counter (modulo $d$) serves this exact purpose. Qutrit measurement also enables error mitigation by detecting accidental leakages outside the qubit subspace \cite{rosenblum2018fault}. Beyond the near-term era, function evaluation oracles are ubiquitous and can be sped up via a counter. For example, recent work demonstrated that just a single qudit, acting as parity check, can implement an oracle-based quantum algorithm \cite{gedik2015computational}.

\begin{figure}
    \centering
    \includegraphics[width=0.5\textwidth]{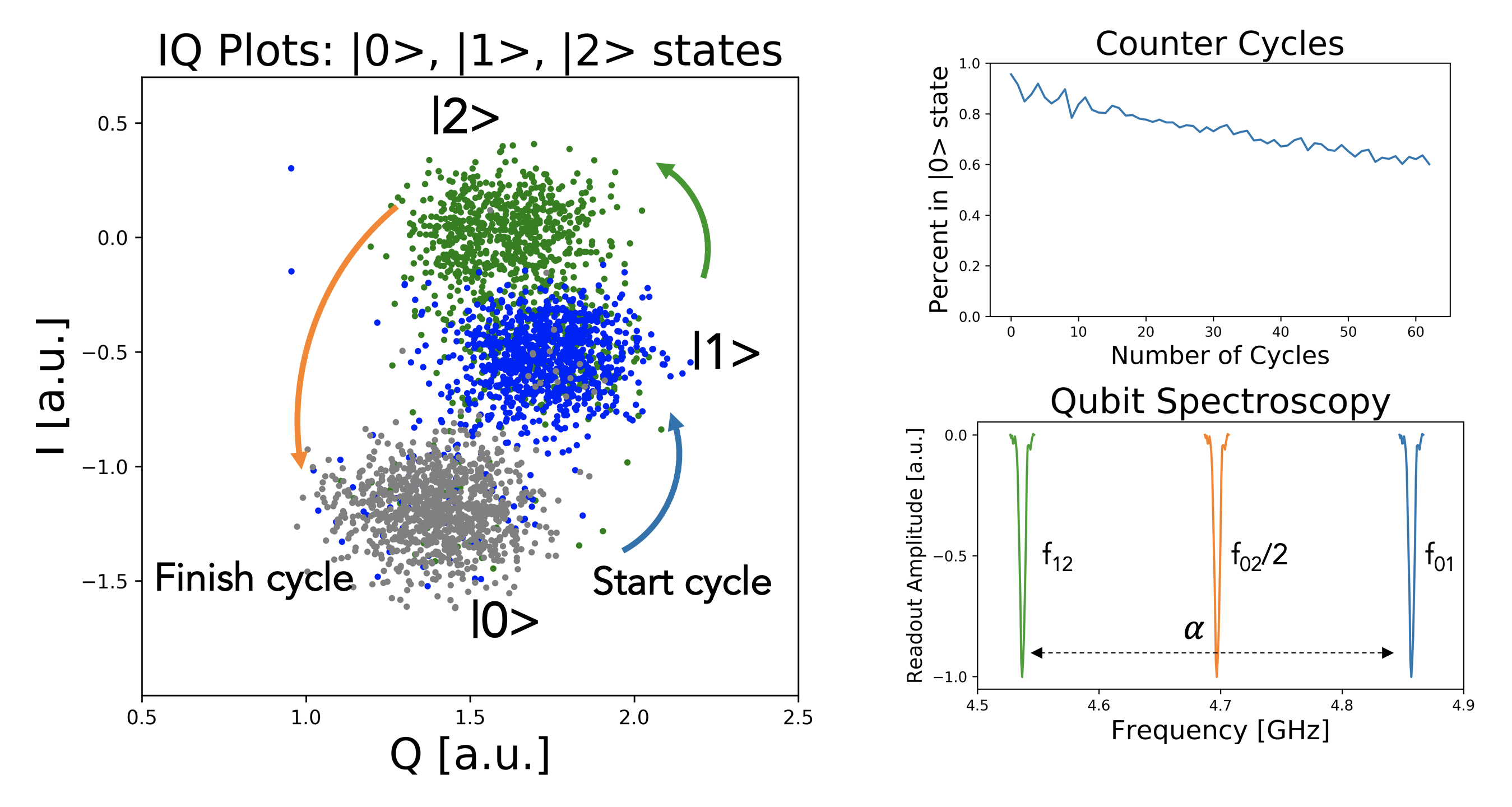}
    \caption{(Left Panel) IQ Plot of readout resonator for different quantum states, and the specific cycle we follow. (Top Right) Percentage of shots found in the ground state as a function of the number of cycles. (Bottom Right) Different transition frequencies for the first three energy levels of a superconducting qubit with $2\pi f_{01}\sim 5GHz$ and $\alpha \sim$-300MHz and a two photon transition. These results span 150k experimental shots on IBM Almaden.}
    \label{counter}
\end{figure}

Here we demonstrate the ability to implement a counter via microwave control of a superconducting qubit, using two transitions previously inaccessible by standard basis gates. Specifically, we target the $f_{12}$ and $f_{02}/2$ transitions (bottom right panel of Figure~\ref{counter}) which act on the $\ket{1}$ to $\ket{2}$ subspace and the $\ket{0}$ to $\ket{2}$ subspace respectively. The required drive strength and duration for these different transitions are dictated by the inherent coupling between each of the levels of interest, which is determined by the physics of the device. In the case of the two photon $f_{02}/2$ transition, the coupling between the $\ket{0}$ and $\ket{2}$ states is suppressed and thereby requires larger drive powers than those needed for an X gate between $\ket{0}\rightarrow \ket{1}$ transitions, with single photon powers around $p_{\text{one}}\approx 0.109$a.u. and two photon powers of $p_{\text{two}}\approx 0.44$a.u., each 35ns in duration. The $f_{12}$ frequency can be measured either by applying an X gate on the $\ket{0}\rightarrow \ket{1}$ transition and subsequently performing qubit spectroscopy, or by driving a two photon $f_{02}/2$ transition and using the prior knowledge of $f_{01}$ to determine $f_{12}$. Once the transition frequencies are identified, we calibrate the proper amplitude and duration of the pulses to fully switch the qubit to the desired final state. 

To gauge the fidelity of our counter, we start off by training a linear discriminator to identify the qutrit state upon readout. In the case of this work, we train a sklearn \cite{scikit-learn} Linear Discriminant Analysis classifier with the calibrated qutrit $\ket{0},\ket{1},\ket{2}$ states and corresponding resonator IQ values (left panel of Figure~\ref{counter}).  Once these calibrations are made, we measure the percentage of shots that have the qutrit in the $\ket{0}$ state at the end of the cycle. Due to imperfections in microwave control, our results deviate from the ideal of 1.0 as the number of cycles increase, making this an ideal testbed for further research such as improved microwave control \cite{kabytayev2014robustness,edmunds2019dynamically} and optimal readout parameters \cite{walter2017rapid}.  Nonetheless, the results indicate remarkably high fidelity---we can drive 60 cycles or 180 hops, before ``dropout'' exceeds 40\%. This appears promising for counting or parity check applications.

\section{Results and Discussion} \label{sec:results}
 \subsection{Benchmarks}
We applied our compiler towards full quantum algorithms. Before proceeding, we note two thematic differences between our treatment of experimental benchmarks and that done in recent architectural work.

First, we focus exclusively on near-term algorithms. Some recent work \cite{murali2019full, murali2019noise, das2019case, tannu2019not, linke2017experimental} demonstrated impressive compiler optimizations for algorithms like Bernstein-Vazirani \cite{bernstein1997quantum}, Hidden-Shift \cite{childs2007quantum}, Adders, and Quantum Fourier Transform \cite{coppersmith2002approximate}. However, we emphasize that these algorithms are not representative of near-term algorithms, which are generally based on a \textit{Hamiltonian simulation} kernel that quantum computers can naturally compute efficiently. Hamiltonian simulation, and thus near-term algorithms broadly, are dominated by the ZZ Interaction optimized in Section~\ref{sec:cross_resonance_optimizations}. We specifically evaluated three types of near-term algorithms: (1) Variational Quantum Eigensolver (VQE) \cite{peruzzo2014variational}, which addresses minimum-eigenvalue problems such as molecular ground state estimation; (2) Quantum Approximate Optimization Algorithm (QAOA) \cite{farhi2014quantum}, which approximates solutions to NP-Hard combinatorial optimization problems; and (3) Hamiltonian Dynamics, which models molecular dynamics and was recently adapted for near-term applications \cite{grimsley2019adaptive, otten2019noise}.

Second, we use Hellinger error/distance (or its complement, Hellinger fidelity) as our top-level metric. Intuitively, Hellinger error captures the distance between two probability distributions: two identical distributions have the minimum distance of 0 and two completely antipodal distributions have the maximum distance of 1. Often, it is appealing to use Probability of Success (i.e. of finding the MAXCUT) as the top-level metric for algorithms like QAOA-MAXCUT \cite{tannu2019ensemble, tannu2019mitigating}. However, QAOA is not intended to find the MAXCUT with 100\% successful probability (otherwise it would solve NP-hard problems in polynomial time), so a QAOA experiment with 100\% ``Probability of Success'' would actually reflect high error. Instead, QAOA is intended to compute a \textit{distribution} of measurement outcomes, within which bitstrings with large cuts will have boosted probabilities. This motivates our use of Hellinger error, and we urge subsequent experimental work to also evaluate near-term algorithms on the basis of probability-distribution distances.

 \subsection{Results}
\begin{figure}
    \centering
    \includegraphics[width=0.48\textwidth]{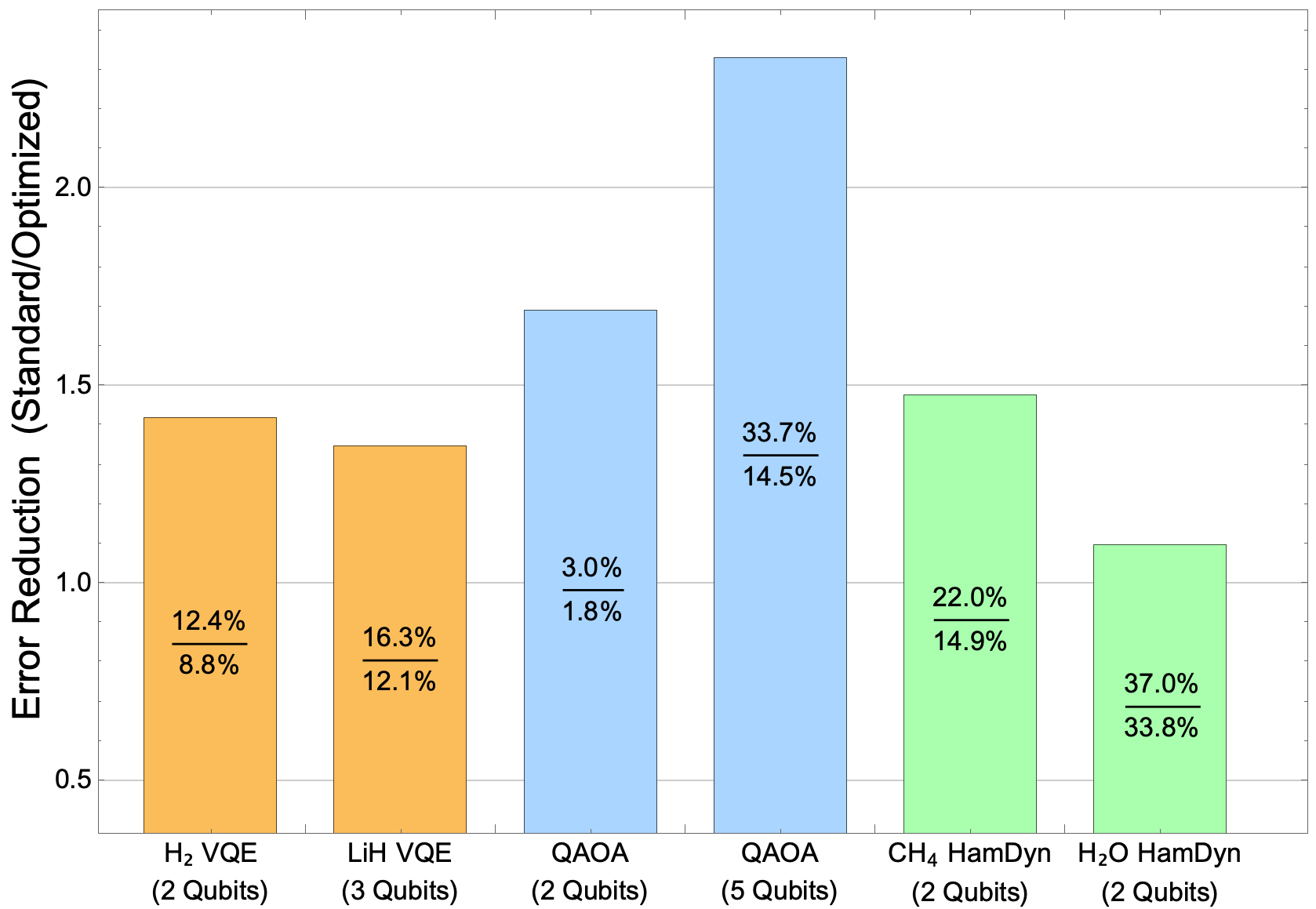} 
    \caption{Reduction in error (Hellinger distance) for benchmarks, due to our optimizations. These results reflect $6 \times 2 \times 8000 = $ 96k shots on IBM Almaden.}
    \label{fig:results}
\end{figure}

Figure~\ref{fig:results} shows the reduction in error due to our optimizations. The H\textsubscript{2} and LiH VQE benchmarks replicate recent experimental work, \cite{o2016scalable} and \cite{hempel2018quantum} respectively. Both experiments are based on the Unitary Coupled Cluster ansatz \cite{romero2018strategies}. The QAOA benchmarks compute MAXCUT on an N-qubit line graph. The Hamiltonian dynamics simulation benchmarks both simulate 6 Trotter steps. The methane and water Hamiltonians were generated with OpenFermion \cite{mcclean2017openfermion}, taking advantage of orbital reductions to reduce the problems to two qubits.

For all six benchmarks, our optimized programs run with much lower error (Hellinger distance/infidelity) between the actual and target outcome distributions. The average error reduction factor is 1.55x and the largest benchmark, 5 Qubit QAOA, has a 2.32x reduction in error from 33.7\% to 14.5\%. The majority of our error reduction stems from our optimization of the ZZ Interaction by augmenting the basis gates with direct access to the Cross-Resonance pulse. We focus on Hellinger error because it is accepted in the quantum community as an (in)fidelity metric and has a ``linear'' interpretation. The average 1.55x error reduction factor is comparable to a year worth of hardware progress; of course, our method is achievable now and is performed in software. Similar work for QAOA was also recently demonstrated on Rigetti's hardware, using a parametrized ZX interaction \cite{abrams2019implementation}.

In addition to the six qubit benchmarks, we also ran the qutrit incrementer in Section~\ref{sec:qudit_operations} and demonstrated 60 cycles, i.e. 180 increment operations, before ``dropout'' exceeds 40\%. This benchmark is unique, because it has no standard qubit comparison---a single qubit cannot model a base-3 counter. This high-fidelity qutrit control confirms that pulse-backed basis gates offer a promising path towards qudit-based optimizations.

\subsection{Source of Fidelity Improvements} \label{subsec:improvement_source}
The fidelity improvements presented here have three sources:
\begin{enumerate}
    \item \textbf{Shorter pulses}. Our compiler's optimized pulses are shorter: 2x shorter for the single qubit rotations in Optimization 1, 24\% shorter for open-CNOTs due to Optimization 2, and $\sim$2x shorter for ZZ interactions due to Optimization 3. These lower operation latencies are advantageous because qubits have less time to decohere.
    \item \textbf{Less calibration error susceptibility}. \texttt{DirectRx($\theta$)} only applies one pre-calibrated (and then amplitude-downscaled) pulse. By contrast, the standard decomposition applies \textit{two} pre-calibrated pulses, squaring the impact of calibration imperfections.
    \item \textbf{Smaller pulse amplitudes}. Our pulse shaping techniques either vertically downscale amplitudes (Optimization 1) or horizontally stretch pulses (Optimization 3). As such, our pulse amplitudes are smaller than or equal to those generated by standard compilation. This is beneficial because smaller pulse amplitudes have smaller spectral components, reducing leakage to undesired frequency sidebands---see Figure 14 in \cite{krantz2019quantum} for details.
\end{enumerate}

Our experience indicates that all three of these sources have meaningful contribution to the fidelity improvements. To further understand our fidelity improvements and reduce the impact of State Preparation and Measurement errors, we performed a Randomized Benchmarking \cite{knill2008randomized} style experiment. In the experiment, we select $K-1$ random single-qubit unitary operations. We execute these $K-1$ operations, terminated with 1 final single-qubit operation that inverts all of the preceding operations. Therefore, under noise-free execution, the qubit returns to the initial state of $\ket{0}$ with 100\% probability. However, due to noise, error accumulates as we increase $K$ from 2 to 25.

\begin{figure}
    \centering
    \includegraphics[width=0.48\textwidth]{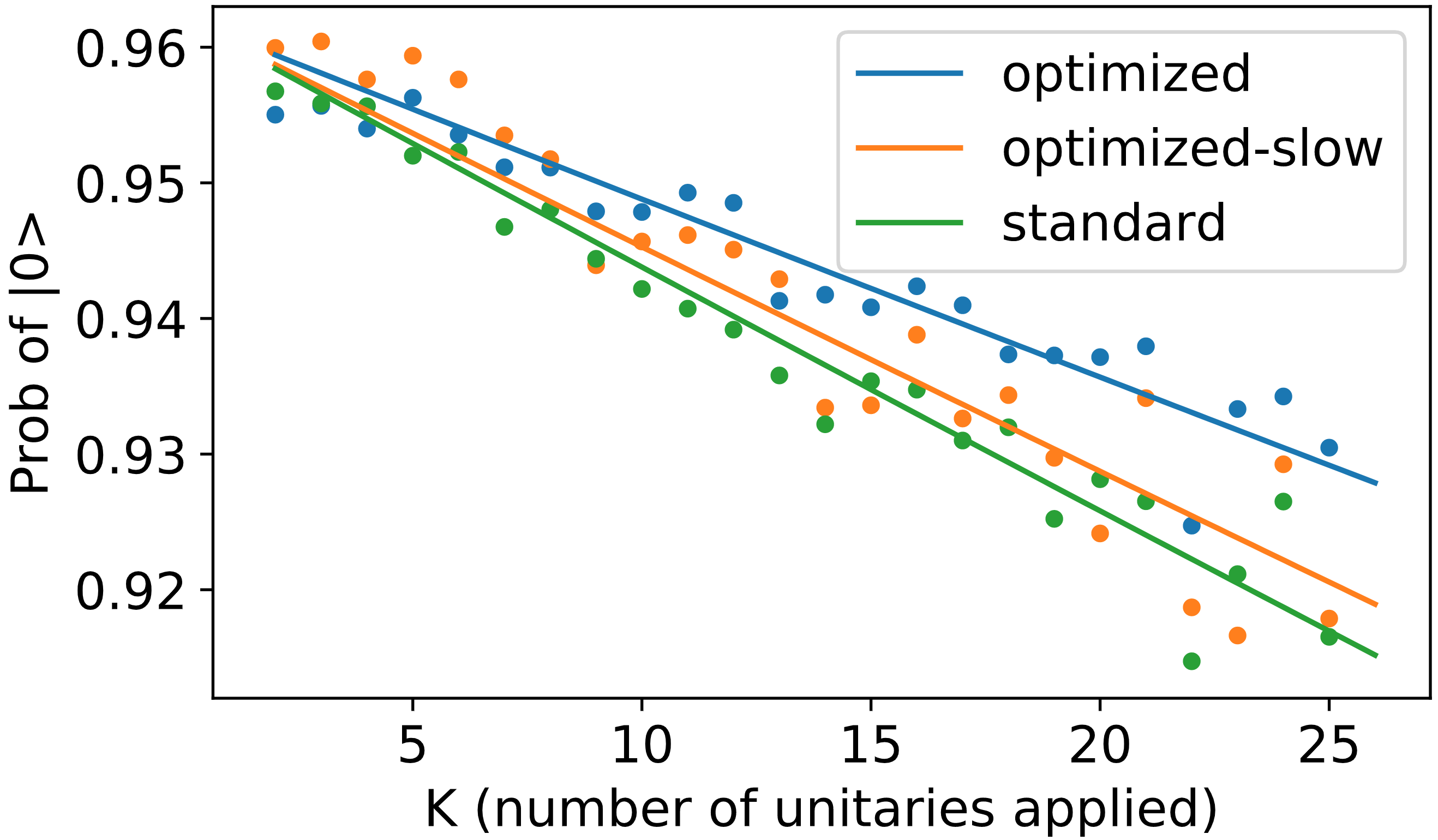} 
    \caption{Randomized Benchmarking style experiment, fit to exponential decay. For each $K$, we randomized 5 sequences of unitary operations. $5 \times 24 \times 3 \times 8$k = 2.88M total shots.}
    \label{fig:rb}
\end{figure}

Figure~\ref{fig:rb} presents our results, which ran over several hours on IBM's Armonk device. The \textit{optimized} plot results from compiling with Optimization 1: Direct Rotations. However, to isolate the effect of shorter pulses, we also compiled \textit{optimized-slow}, which inserts NO-OP idling into the optimized pulse schedules, to match the duration of the standard pulse schedules.

Each trajectory was fit to the exponential decay, $f^K - b$, where $b$ is a y-intercept term that represents SPAM errors independent of $K$, and $f$ is interpreted as gate fidelity. The resulting gate fidelities for optimized, optimized-slow, and standard are $f$ = 99.87\%, 99.83\%, and 99.82\%. This implies that shorter pulses (\#1) account for 70\% of the fidelity improvement, while less susceptibility to calibration imperfection (\#2) and smaller pulse amplitudes (\#3) account for the remaining 30\% improvement. The improvement due to shorter pulses matches theoretical predictions: according to the \textit{gate error in coherence limit} calculation \cite[Eq. 24]{naik2017random}, the 2x pulse speedup yields a minimum 0.01\% fidelity improvement.

\section{Conclusion} \label{sec:conclusion}

Our results demonstrate that augmenting basis gates with pulse backed hardware primitives, bootstrapped from existing calibrations, leads to 1.6x error reductions and 2x speedups for near-term algorithms. Critically, our technique does not rely on knowledge of the system Hamiltonian, thus bypassing the experimental barriers to quantum optimal control. The measured fidelity improvements are arguably equivalent to a year's worth of hardware progress, but our techniques are available immediately, through software. We hope that our experiences with OpenPulse will encourage more quantum vendors to expose their hardware to pulse-level control. To this end, all of our code and notebooks are available on Github \cite{anonymized_repo}.

%\appendix
%\input{sections/10appendix.tex}

\section*{Acknowledgements}
This work is funded in part by EPiQC, an NSF Expedition in Computing, under grants CCF-1730449/1832377; in part by STAQ under grant NSF Phy-1818914; and in part by DOE grants DE-SC0020289 and DE-SC0020331. This research used resources of the Oak Ridge Leadership Computing Facility, which is a DOE Office of Science User Facility supported under Contract DE-AC05-00OR22725. P. G. is supported by the Department of Defense (DoD) through the National Defense Science \& Engineering Graduate Fellowship (NDSEG) Program.

%%%%%%% -- PAPER CONTENT ENDS -- %%%%%%%%

%%%%%%%%% -- BIB STYLE AND FILE -- %%%%%%%%
\bibliographystyle{ieeetr}
\bibliography{refs}
%%%%%%%%%%%%%%%%%%%%%%%%%%%%%%%%%%%%

\end{document}